\def\year{2019}\relax
\documentclass[letterpaper]{article} 
\pdfoutput=1
\usepackage{aaai19}  
\usepackage{times}  
\usepackage{helvet}  
\usepackage{courier}  
\usepackage{url}  
\usepackage{graphicx}  
\newif\ifArxivVersion
\ArxivVersionfalse
\ArxivVersiontrue  

\usepackage[utf8]{inputenc} 
\usepackage[T1]{fontenc}    
\usepackage{url}            
\usepackage{booktabs}       
\usepackage{amsfonts}       
\usepackage{nicefrac}       
\usepackage{microtype}      
\usepackage{graphicx}
\usepackage{xcolor}
\usepackage[scaled=.8]{beramono} 
\usepackage{amsmath}
\usepackage{amsthm}
\usepackage{amsfonts}
\usepackage{amssymb}
\usepackage{bm}
\usepackage{array}
\usepackage{xspace}
\usepackage{mathtools}
\usepackage{subcaption}
\usepackage{algorithm}
\usepackage{dsfont}
\usepackage{fontawesome}
\usepackage{thmtools}
\usepackage{thm-restate}
\usepackage[]{algpseudocode}
\usepackage{xspace}
\usepackage{lipsum}
\usepackage{pifont}
\usepackage{makecell}
\usepackage{tablefootnote}
\usepackage[tight-spacing=true]{siunitx}
\usepackage{wrapfig}
\usepackage{multirow}
\usepackage{afterpage}
\usepackage{tabularx} 

\usepackage[capitalize]{cleveref} 
\crefname{equation}{}{} 

\newcommand{\be}{\begin{enumerate}}
\newcommand{\ee}{\end{enumerate}}
\newcommand{\bi}{\begin{itemize}}
\newcommand{\ei}{\end{itemize}}

\newcommand{\meta}[1]{\widetilde{#1}}
\newcommand{\joint}[1]{\bm{#1}}
\newcommand{\jm}[1]{\joint{\meta{#1}}}

\newcommand{\header}[1]{\textbf{#1}\,\,}

\newcommand{\paperTitle}{Learning to Teach in Cooperative Multiagent Reinforcement Learning} 

\usepackage{anyfontsize} 

\newcommand{\Ptask}{$\mathcal{P}_{\text{Task}}$\xspace}
\newcommand{\Padvise}{$\meta{\mathcal{P}}_{\text{Advise}}$\xspace}

\newcommand{\inlineask}{S} 
\newcommand{\inlineteach}{T} 
\newcommand{\ask}{S} 
\newcommand{\teach}{T} 

\newtheorem{remark}{Remark}

\newif\ifcomments
\commentsfalse 
\commentstrue 

\ifcomments
	\newcommand{\sXX}[1]{\color{red}SO: (#1)\color{black}\xspace}
	\newcommand{\dXX}[1]{\color{olive}DK: (#1)\color{black}\xspace}
	\newcommand{\jXX}[1]{\color{orange}JH: (#1)\color{black}\xspace}
	\newcommand{\mXX}[1]{\color{cyan}ML: (#1)\color{black}\xspace}
	\newcommand{\gXX}[1]{\color{purple}GT: (#1)\color{black}\xspace}
	\newcommand{\mrXX}[1]{\color{blue}MR: (#1)\color{black}}
	\newcommand{\cXX}[1]{\color{olive}CA: (#1)\color{black}}
\else
	\newcommand{\sXX}[1]{}
	\newcommand{\dXX}[1]{}
	\newcommand{\jXX}[1]{}
	\newcommand{\mXX}[1]{}
	\newcommand{\gXX}[1]{}
	\newcommand{\mrXX}[1]{}
	\newcommand{\cXX}[1]{}
\fi

\begin{document}
\title{\paperTitle}

\ifArxivVersion
	{ 
		\author{
			\fontsize{10pt}{13pt} \selectfont
			Shayegan Omidshafiei\textsuperscript{{\normalfont1,2}}\\
			\fontsize{9pt}{12pt} \selectfont
			\texttt{shayegan@mit.edu}
			\And
			\fontsize{10pt}{13pt} \selectfont
			Dong-Ki Kim\textsuperscript{{\normalfont1,2}}\\
			\fontsize{9pt}{12pt} \selectfont
			\texttt{dkkim93@mit.edu}
			\And
			\fontsize{10pt}{13pt} \selectfont
			Miao Liu\textsuperscript{{\normalfont2,3}}\\
			\fontsize{9pt}{12pt} \selectfont
			\texttt{miao.liu1@ibm.com}
			\And
			\fontsize{10pt}{13pt} \selectfont
			Gerald Tesauro\textsuperscript{{\normalfont2,3}}\\
			\fontsize{9pt}{12pt} \selectfont
			\texttt{gtesauro@us.ibm.com}
			\AND
			\fontsize{10pt}{13pt} \selectfont
			Matthew Riemer\textsuperscript{{\normalfont2,3}}\\
			\fontsize{9pt}{12pt} \selectfont
			\texttt{mdriemer@us.ibm.com}
			\And
			\fontsize{10pt}{13pt} \selectfont
			Christopher Amato\textsuperscript{{\normalfont4}}\\
			\fontsize{9pt}{12pt} \selectfont			
			\texttt{camato@ccs.neu.edu}
			\And
			\fontsize{10pt}{13pt} \selectfont
			Murray Campbell\textsuperscript{{\normalfont2,3}}\\
			\fontsize{9pt}{12pt} \selectfont			
			\texttt{mcam@us.ibm.com}
			\And
			\fontsize{10pt}{13pt} \selectfont
			Jonathan P.~How\textsuperscript{{\normalfont1,2}}\\
			\fontsize{9pt}{12pt} \selectfont			
			\texttt{jhow@mit.edu}
			\AND
			\normalfont
			\fontsize{9pt}{12pt} \selectfont
			\normalsize\textsuperscript{1}LIDS, MIT \hspace{1em}
			\normalsize\textsuperscript{2}MIT-IBM Watson AI Lab \hspace{1em}
			\normalsize\textsuperscript{3}IBM Research \hspace{1em}
			\normalsize\textsuperscript{4}CCIS, Northeastern University
		}
	}
\else
	\author{Anonymous Authors (Paper ID 2545)}
\fi

\maketitle

\begin{abstract}
	Collective human knowledge has clearly benefited from the fact that innovations by individuals are taught to others through communication. 
	Similar to human social groups, agents in distributed learning systems would likely benefit from communication to share knowledge and teach skills.
	The problem of teaching to improve agent learning has been investigated by prior works, but these approaches make assumptions that prevent application of teaching to general multiagent problems, or require domain expertise for problems they can apply to.
	This learning to teach problem has inherent complexities related to measuring long-term impacts of teaching that compound the standard multiagent coordination challenges.
	In contrast to existing works, this paper presents the first general framework and algorithm for intelligent agents to learn to teach in a multiagent environment. 
	Our algorithm, Learning to Coordinate and Teach Reinforcement (LeCTR), addresses peer-to-peer teaching in cooperative multiagent reinforcement learning. 
	Each agent in our approach learns both when and what to advise, then uses the received advice to improve local learning. 
	Importantly, these roles are not fixed; these agents learn to assume the role of student and/or teacher at the appropriate moments, requesting and providing advice in order to improve teamwide performance and learning.
	Empirical comparisons against state-of-the-art teaching methods show that our teaching agents not only learn significantly faster, but also learn to coordinate in tasks where existing methods fail. 
\end{abstract}

\section{Introduction} \label{sec:intro}
In social settings, innovations by individuals are taught to others in the population through communication channels \cite{rogers2010diffusion}, which not only improves final performance, but also the effectiveness of the entire learning process (i.e., rate of learning).
There exist analogous settings where learning agents interact and adapt behaviors while interacting in a shared environment (e.g., autonomous cars and assistive robots).
While any given agent may not be an expert during learning, it may have local knowledge that teammates may be unaware of.
Similar to human social groups, these learning agents would likely benefit from communication to share knowledge and teach skills, thereby improving the effectiveness of system-wide learning.
It is also desirable for agents in such systems to learn to teach one another, rather than rely on hand-crafted teaching heuristics created by domain experts.
The benefit of learned peer-to-peer teaching is that it can accelerate learning even without relying on the existence of ``all-knowing'' teachers.
Despite these potential advantages, no algorithms exist for learning to teach in multiagent systems.

This paper targets the learning to teach problem in the context of cooperative Multiagent Reinforcement Learning (MARL). Cooperative MARL is a standard framework for settings where agents learn to coordinate in a shared environment.
Recent works in cooperative MARL have shown final task performance can be improved by introducing inter-agent communication mechanisms \cite{sukhbaatar2016learning,foerster2016learning,lowe2017multi}.
Agents in these works, however, merely communicate to coordinate in the given task, not to improve overall learning by teaching one another. 
By contrast, this paper targets a new multiagent paradigm in which agents learn to teach by communicating action advice, thereby improving final performance and accelerating teamwide learning.

The learning to teach in MARL problem has unique inherent complexities that compound the delayed reward, credit assignment, and partial observability issues found in general multiagent problems \cite{oliehoek2016concise}.
As such, there are several key issues that must be addressed. 
First, agents must learn when to teach, what to teach, and how to learn from what is being taught.
Second, despite coordinating in a shared environment, agents may be independent/decentralized learners with privacy constraints (e.g., robots from distinct corporations that cannot share full policies), and so must learn how to teach under these constraints. 
A third issue is that agents must estimate the impact of each piece of advice on their teammate's learning progress.
Delays in the accumulation of knowledge make this credit assignment problem difficult, even in supervised/unsupervised learning \cite{graves2017automated}. 
Nonstationarities due to agent interactions and the temporally-extended nature of MARL compound these difficulties in our setting.
These issues are unique to our learning to teach setting and remain largely unaddressed in the literature, despite being of practical importance for future decision-making systems.  
One of the main reasons for the lack of progress addressing these inherent challenges is the significant increase in the computational complexity of this new teaching/learning paradigm compared to multiagent problems that have previously been considered.

Our paper targets the problem of learning to teach in a multiagent team, which has not been considered before.
Each agent in our approach learns both when and what to advise, then uses the received advice to improve local learning. 
Importantly, these roles are not fixed (see \cref{fig:robots_lectr_overview}); these agents learn to assume the role of student and/or teacher at appropriate moments, requesting and providing advice to improve teamwide performance and learning.
In contrast to prior works, our algorithm supports teaching of heterogeneous teammates and applies to settings where advice exchange incurs a communication cost.
Comparisons conducted against state-of-the-art teaching methods show that our teaching agents not only learn significantly faster, but also learn to coordinate in tasks where existing methods fail.
\section{Background: Cooperative MARL} \label{sec:background}
Our work targets cooperative MARL, where agents execute actions that jointly affect the environment, then receive feedback via local observations and a shared reward.
This setting is formalized as a Decentralized Partially Observable Markov Decision Process (Dec-POMDP),  defined as $\langle \mathcal{I}, \mathcal{S}, \bm{\mathcal{A}}, \mathcal{T}, \mathcal{R}, \bm{\Omega}, \mathcal{O}, \gamma \rangle$ \cite{oliehoek2016concise}; 
$\mathcal{I}$ is the set of $n$ agents, $\mathcal{S}$ is the state space, $\bm{\mathcal{A}} = \times_{i} \mathcal{A}^{i}$ is the joint action space, and $\bm{\Omega} = \times_{i} \Omega^{i}$ is the joint observation space.\footnote{Superscript $i$ denotes parameters for the $i$-th agent. Refer to the supplementary material for a notation list.} 
Joint action $\bm{a} = \langle a^{1}, \ldots, a^{n} \rangle$ causes state $s \in \mathcal{S}$ to transition to $s' \in \mathcal{S}$ with probability $P(s'|s,\bm{a}) = \mathcal{T}(s,\bm{a},s')$. 
At each timestep $t$, joint observation $\bm{o} = \langle o^{1}, \ldots, o^{n} \rangle$ is observed with probability $P(\bm{o}|s',\bm{a}) = \mathcal{O}(\bm{o},s',\bm{a})$.
Given its observation history, $h_t^{i} = (o^{i}_{1},\ldots,o^{i}_{t})$, agent $i$ executes actions dictated by its policy $a^{i}=\pi^{i}(h_{t}^{i})$. 
The joint policy is denoted by $\bm{\pi} = \langle \pi^{1}, \ldots, \pi^{n} \rangle$ and parameterized by $\joint{\theta}$. 
It may sometimes be desirable to use a recurrent policy representation (e.g., recurrent neural network) to compute an internal state $h_t$ that compresses the observation history, or to explicitly compute a belief state (probability distribution over states); with abuse of notation, we use $h_t$ to refer to all such variations of internal states/observation histories.
At each timestep, the team receives reward $r_t = \mathcal{R}(s_t,\bm{a}_t)$, with the objective being to maximize value, $V(s;\joint{\theta}) = \mathbb{E}[\sum_{t}\gamma^{t}r_t|s_0=s]$, given discount factor $\gamma \in [0,1)$. 
Let action-value $Q^{i}(o^{i},a^{i};h^{i})$ denote agent $i$'s expected value for executing action $a^{i}$ given a new local observation $o^{i}$ and internal state $h^{i}$, and using its policy thereafter.
We denote by $\vec{Q}(o^{i};h^{i})$ the vector of action-values (for all actions) given new observation $o^{i}$.

\section{Teaching in Cooperative MARL}
\begin{figure}[t]
	\centering
	\includegraphics[width=0.95\linewidth]{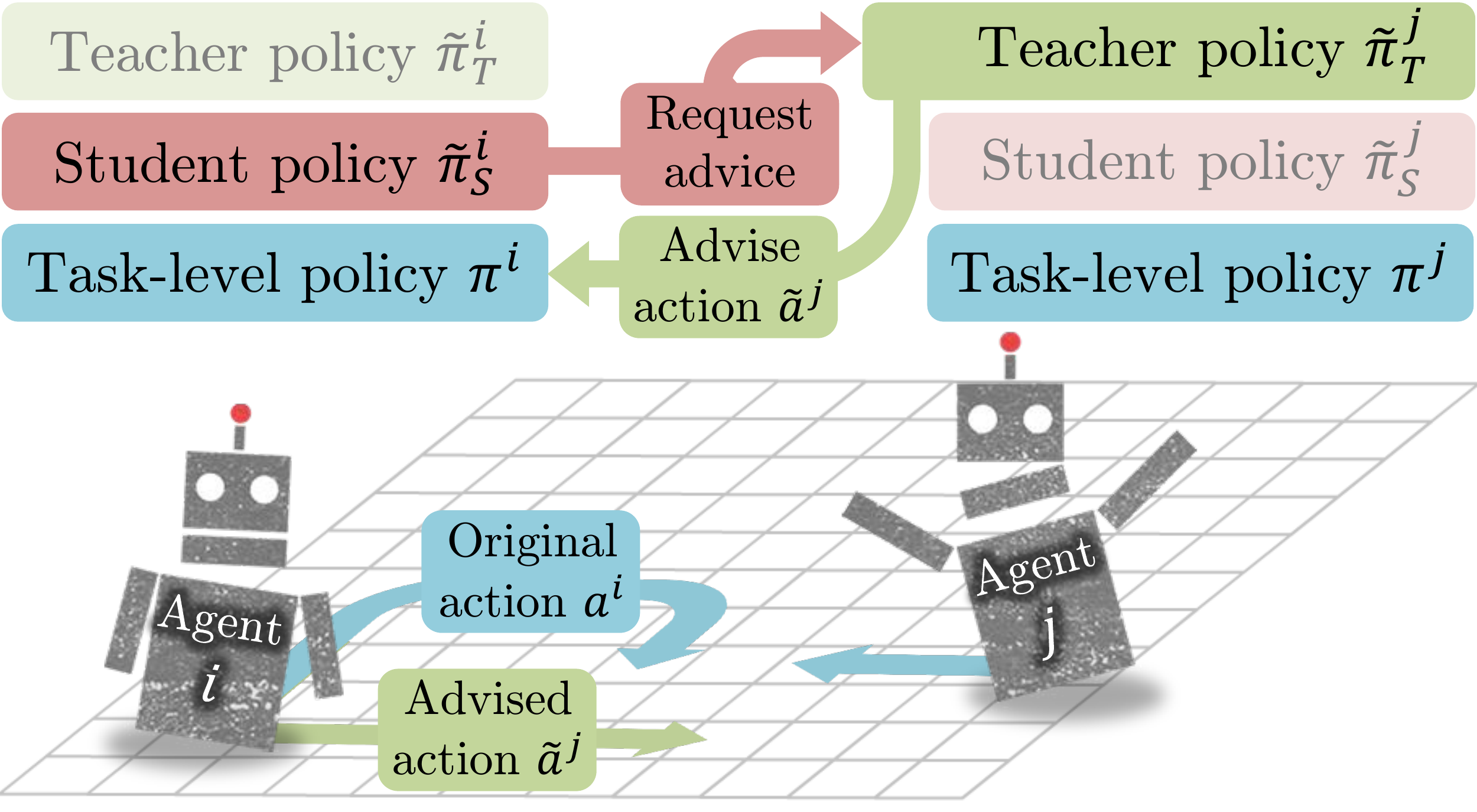}
	\caption{Overview of teaching via action advising in MARL. Each agent learns to execute the task using task-level policy $\pi$, to request advice using learned student policy $\meta{\pi}_{\ask}$, and to respond with action advice using learned teacher policy $\meta{\pi}_{\teach}$. Each agent can assume a student and/or teacher role at any time. In this example, agent $i$ uses its student policy to request help, agent $j$ advises action $\meta{a}^{j}$, which the student executes instead of its originally-intended action $a^{i}$. By learning to transform the local knowledge captured in task-level policies into action advice, the agents can help one another learn.}
	\label{fig:robots_lectr_overview}
\end{figure}

This work explores multiagent teaching in a setting where no agent is necessarily an all-knowing expert.
This section provides a high-level overview of the motivating scenario.
Consider the cooperative MARL setting in \cref{fig:robots_lectr_overview}, where agents $i$ and $j$ learn a joint task (i.e., a Dec-POMDP).
In each learning iteration, these agents interact with the environment and collect data used by their respective learning algorithms, $\mathbb{L}^{i}$ and $\mathbb{L}^{j}$, to update their policy parameters, $\theta^{i}$ and $\theta^{j}$. 
This is the standard cooperative MARL problem, which we hereafter refer to as \Ptask: \textbf{the task-level learning problem}.
For example, task-level policy $\pi^{i}$ is the policy agent $i$ learns and uses to execute actions in the task.
Thus, task-level policies summarize each agent's learned behavioral knowledge.

During task-level learning, it is unlikely for any agent to be an expert.
However, each agent may have unique experiences, skill sets, or local knowledge of how to learn effectively in the task.
Throughout the learning process, it would likely be useful for agents to advise one another using this local knowledge, in order to improve final performance and accelerate teamwide learning. 
Moreover, it would be desirable for agents to learn when and what to advise, rather than rely on hand-crafted and domain-specific advising heuristics.
Finally, following advising, agents should ideally have learned effective task-level policies that no longer rely on teammate advice at every timestep.
We refer to this new problem, which involves agents learning to advise one another to improve joint task-level learning, as \Padvise: \textbf{the advising-level problem}.

The advising mechanism used in this paper is \emph{action advising}, where agents suggest actions to one another. 
By learning to appropriately transform local knowledge (i.e., task-level policies) into action advice, teachers can affect students' experiences and their resulting task-level policy updates.
Action advising makes few assumptions, in that learners need only use task-level algorithms $\langle \mathbb{L}^{i}, \mathbb{L}^{j} \rangle$ that support off-policy exploration (enabling execution of action advice for policy updates), and that they receive advising-level observations summarizing teammates' learning progress (enabling learning of when/what to advise). 
Action advising has a good empirical track record \cite{torrey2013teaching,taylor2014reinforcement,fachantidis2017learning,da2017simultaneously}.
However, existing frameworks have key limitations:
the majority are designed for single-agent RL and do not consider multiagent learning; 
their teachers always advise optimal actions to students, making decisions about when (not what) to teach; 
they also use heuristics for advising, rather than training teachers by measuring student learning progress.
By contrast, agents in our paper learn to interchangeably assume the role of a student (advice requester) and/or teacher (advice responder), denoted $\inlineask$ and $\inlineteach$, respectively.
Each agent learns task-level policy $\pi$ used to actually perform the task, student policy $\meta{\pi}_{\ask}$ used to request advice during task-level learning, and teacher policy $\meta{\pi}_{\teach}$ used to advise a teammate during task-level learning.\footnote{Tilde accents (e.g., $\meta{\pi}$) denote advising-level properties.}

Before detailing the algorithm, let us first illustrate the multiagent interactions in this action advising scenario.
Consider again \cref{fig:robots_lectr_overview}, where agents are learning to execute a task (i.e., solving \Ptask) while advising one another.
While each agent in our framework can assume a student and/or teacher role at any time, \cref{fig:robots_lectr_overview} visualizes the case where agent $i$ is the student and agent $j$ the teacher.
At a given task-level learning timestep, agent $i$'s task-level policy $\pi^{i}$ outputs an action (`original action $a^{i}$' in \cref{fig:robots_lectr_overview}).
However, as the agents are still learning to solve \Ptask, agent $i$ may prefer to execute an action that maximizes local learning. 
Thus, agent $i$ uses its student policy $\meta{\pi}_{\ask}^{i}$ to decide whether to ask teammate $j$ for advice. 
If this advice request is made, teammate $j$ checks its teacher policy $\meta{\pi}_{\teach}^{j}$ and task-level policy $\pi^{j}$ to decide whether to respond with action advice.
Given a response, agent $i$ then executes the advised action ($\meta{a}^{j}$ in \cref{fig:robots_lectr_overview}) as opposed to its originally-intended action ($a^{i}$ in \cref{fig:robots_lectr_overview}). 
This results in a local experience that agent $i$ uses to update its task-level policy.
A reciprocal process occurs when the agents' roles are reversed.
The benefit of advising is that agents can learn to use local knowledge to improve teamwide learning.

Similar to recent works that model the multiagent learning process \cite{hadfield2016cooperative,foerster2018learning}, we focus on the pairwise (two agent) case, targeting the issues of when/what to advise, then detail extensions to $n$ agents.
Even in the pairwise case, there exist issues unique to our learning to teach paradigm.
First, note that the objectives of \Ptask and \Padvise are distinct.
Task-level problem, \Ptask, has a standard MARL objective of agents learning to coordinate to maximize final performance in the task. 
Learning to advise (\Padvise), however, is a higher-level problem, where agents learn to influence teammates' task-level learning by advising them.
However, \Ptask and \Padvise are also coupled, as advising influences the task-level policies learned.
Agents in our problem must learn to advise despite the nonstationarities due to changing task-level policies, which are also a function of algorithms $\langle \mathbb{L}^{i}, \mathbb{L}^{j} \rangle$ and policy parameterizations $\langle \theta^{i}, \theta^{j} \rangle$.

Learning to teach is also distinct from prior works that involve agents learning to communicate \cite{sukhbaatar2016learning,foerster2016learning,lowe2017multi}. 
These works focus on agents communicating in order to coordinate in a task.
By contrast, our problem focuses on agents learning how advising affects the underlying task-level learning process, then using this knowledge to accelerate learning even when agents are non-experts.
Thus, the objectives of communication-based multiagent papers are disparate from ours, and the two approaches may even be combined.
\section{LeCTR: Algorithm for Learning to Coordinate and Teach Reinforcement}\label{sec:lectr_algorithm}
This section introduces our learning to teach approach, details how issues specific to our problem setting are resolved, and summarizes overall training protocol.
Pseudocode is presented in the supplementary material due to limited space.

\begin{figure}[t]
	\centering
	\includegraphics[width=0.95\linewidth,page=]{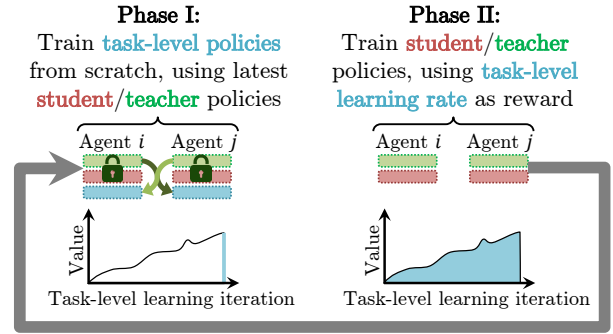}
	\caption{LeCTR consists of two iterated phases: task-level learning (Phase I), and advising-level learning (Phase II). In Phase II, advising policies are trained using rewards correlated to task-level learning (see \cref{table:metareward_summary}). Task-level, student, and teacher policy colors above follows convention of \cref{fig:robots_lectr_overview}.}
	\label{fig:training_protocol}
\end{figure}

\header{Overview} 
Our algorithm, Learning to Coordinate and Teach Reinforcement (LeCTR), solves advising-level problem \Padvise. 
The objective is to learn advising policies that augment agents' task-level algorithms $\langle \mathbb{L}^{i}, \mathbb{L}^{j} \rangle$ to accelerate solving of \Ptask. 
Our approach involves 2 phases (see \cref{fig:training_protocol}): 
\begin{itemize}
	\item Phase I: agents learn \Ptask from scratch using blackbox learning algorithms $\langle \mathbb{L}^{i}, \mathbb{L}^{j} \rangle$ and latest advising policies.
	\item Phase II: advising policies are updated using advising-level rewards correlated to teammates' task-level learning.
\end{itemize}
No restrictions are placed on agents' task-level algorithms (i.e., they can be heterogeneous).
Iteration of Phases I and II enables training of increasingly capable advising policies. 

\header{Advising Policy Inputs \& Outputs} 
LeCTR learns student policies $\langle \meta{\pi}^{i}_{\ask}, \meta{\pi}^{j}_{\ask} \rangle$ and teacher policies $\langle \meta{\pi}^{i}_{\teach}, \meta{\pi}^{j}_{\teach} \rangle$ for agents $i$ and $j$, constituting a jointly-initiated advising approach that learns {when} to request advice and {when}/{what} to advise.
It is often infeasible to learn high-level policies that directly map task-level policy parameters $\langle\theta^{i},\theta^{j}\rangle$ (i.e., local knowledge) to advising decisions: 
the agents may be independent/decentralized learners and the cost of communicating task-level policy parameters may be high; 
sharing policy parameters may be undesirable due to privacy concerns; 
and learning advising policies over the task-level policy parameter space may be infeasible (e.g., if the latter policies involve millions of parameters).
Instead, each LeCTR agent learns advising policies over advising-level observations $\meta{o}$.
As detailed below, these observations are selected to provide information about agents' task-level state and knowledge in a more compact manner than full policy parameters $\langle\theta^{i},\theta^{j}\rangle$. 

Each LeCTR agent can be a student, teacher, or both simultaneously (i.e., request advice for its own state, while advising a teammate in a different state).
For clarity, we detail advising protocols when agents $i$ and $j$ are student and teacher, respectively (see \cref{fig:robots_lectr_overview}).
LeCTR uses distinct advising-level observations for student and teacher policies.
Student policy $\meta{\pi}^{i}_{\ask}$ for agent $i$ decides when to request advice using advising-level observation $\meta{o}_{\ask}^{i} = \langle o^{i}, \vec{Q}^{i}(o^{i};h^{i}) \rangle$, where $o^{i}$ and $\vec{Q}^{i}(o^{i};h^{i})$ are the agent's task-level observation and action-value vectors, respectively. 
Through $\meta{o}_{\ask}^{i}$, agent $i$ observes a measure of its local task-level observation and policy state.
Thus, agent $i$'s student-perspective action is $\meta{a}^{i}_{\ask}=\meta{\pi}^{i}_{\ask}(\meta{o}_{\ask}^{i}) \in \{\text{request advice},\text{do not request advice}\}$. 

Similarly, agent $j$'s teacher policy $\meta{\pi}^{j}_{\teach}$ uses advising-level observation $\meta{o}_{\teach}^{j} = \langle o^{i}, \vec{Q}^{i}(o^{i};h^{i}), \vec{Q}^{j}(o^{i};h^{i}) \rangle$ to decide when/what to advise. 
$\meta{o}_{\teach}^{j}$ provides teacher agent $j$ with a measure of student $i$'s task-level state/knowledge (via $o^{i}$ and $\vec{Q}^{i}(o^{i};h^{i})$) and of its own task-level knowledge given the student's context (via $\vec{Q}^{j}(o^{i};h^{i})$). 
Using $\meta{\pi}^{j}_{\teach}$, teacher $j$ decides {what} to advise: either an action from student $i$'s action space, $\mathcal{A}^{i}$, or a special no-advice action $\meta{a}_\emptyset$. 
Thus, the teacher-perspective action for agent $j$ is $\meta{a}^{j}_{\teach}=\meta{\pi}^{j}_{\teach}(\meta{o}_{\teach}^{j}) \in \mathcal{A}^{i} \cup \{\meta{a}_\emptyset\}$. 

Given no advice, student $i$ executes originally-intended action $a^{i}$. 
However, given advice $\meta{a}^{j}_{T}$, student $i$ executes action $\beta^{i}(\meta{a}^{j}_{T})$, where $\beta^{i}(\cdot)$ is a local behavioral policy not known by $j$. 
The assumption of local behavioral policies increases the generality of LeCTR, as students may locally transform advised actions before execution.

Following advice execution, agents collect task-level experiences and update their respective task-level policies. 
A key feature is that LeCTR agents learn {what} to advise by training $\langle \meta{\pi}^i_{\teach},\meta{\pi}^j_{\teach}\rangle$, rather than always advising actions they would have taken in students' states.
These agents may learn to advise exploratory actions or even decline to advise if they estimate that such advice will improve teammate learning.


\begin{table*}[t!]
	\caption{Summary of rewards used to train advising policies. Rewards shown are for the case where agent $i$ is student and agent $j$ teacher (i.e., flip the indices for the reverse case). Each reward corresponds to a different measure of task-level learning after the student executes an action advice and uses it to update its task-level policy. Refer to the supplementary material for more details.}
	\begin{tabular}{llc}
		\toprule
		Advising Reward Name        & Description                      &  Reward Value $\meta{r}^{j}_{\teach}=\meta{r}^{i}_{\ask}$   \\ \midrule
		JVG: Joint Value Gain      & Task-level value $V(s;\joint{\theta})$ improvement after learning                           &     $V(s;\joint{\theta}_{t+1})-V(s;\joint{\theta}_{t})$     \\
		QTR: Q-Teaching Reward      & Teacher's estimate of best vs. intended student action                 &      $\max_a Q_{\teach}(\!o^i,a;\!h^i) - Q_{\teach}(\!o^i,a^i;\!h^i)$       \\
		LG: Loss Gain               & Student's task-level loss $\mathcal{L}(\theta^{i})$ reduction                  & $\mathcal{L}(\theta_{t}^{i})-\mathcal{L}(\theta_{t+1}^{i})$ \\
		LGG: Loss Gradient Gain    & Student's task-level policy gradient magnitude                   &    $||\nabla_{\theta^{i}}\mathcal{L}(\theta^{i})||^2_2$     \\
		TDG: TD Gain                & Student's temporal difference (TD) error $\delta^{i}$ reduction   &            $|\delta_{t}^{i}|-|\delta_{t+1}^{i}|$            \\
		VEG: Value Estimation Gain & Student's value estimate $\hat{V}(\theta^{i})$ gain above threshold $\tau$ &           $\mathds{1}(\hat{V}(\theta^{i})>\tau)$            \\ \bottomrule
	\end{tabular}
	\label{table:metareward_summary}
\end{table*}

\header{Rewarding Advising Policies}
Recall in Phase II of LeCTR, advising policies are trained to maximize advising-level rewards that should, ideally, reflect the objective of accelerating task-level learning.
Without loss of generality, we focus again on the case where agents $i$ and $j$ assume student and teacher roles, respectively, to detail these rewards.
Since student policy $\meta{\pi}_{\ask}^{i}$ and teacher policy $\meta{\pi}_{\teach}^{j}$ must coordinate to help student $i$ learn, they receive identical advising-level rewards, $\meta{r}_{\ask}^{i} = \meta{r}_{\teach}^{j}$.
The remaining issue is to identify advising-level rewards that reflect learning progress.
\begin{remark}\label{remark:meta_reward_equals_r_case}
	Earning task-level rewards by executing advised actions may not imply actual learning.
	Thus, rewarding advising-level policies with the task-level reward, $r$, received after advice execution can lead to poor advising policies. 
\end{remark}

We evaluate many choices of advising-level rewards, which are summarized and described in \cref{table:metareward_summary}. 
The unifying intuition is that each reward type corresponds to a different measure of the advised agent's task-level learning, which occurs after executing an advised action. 
Readers are referred to the supplementary material for more details.

Note that at any time, task-level action $a^{i}$ executed by agent $i$ may either be selected by its local task-level policy, or by a teammate $j$ via advising.
In Phase II, pair $\langle \meta{\pi}_{\ask}^{i},\meta{\pi}_{\teach}^{j}\rangle$ is rewarded only if advising occurs (with zero advising reward otherwise). 
Analogous advising-level rewards apply for the reverse student-teacher pairing $j$-$i$, where $\meta{r}_{\ask}^{j} = \meta{r}_{\teach}^{i}$. 
During Phase II, we train all advising-level policies using a joint advising-level reward $\meta{r} = \meta{r}_T^{i} + \meta{r}_T^{j}$ to induce cooperation.

Advising-level rewards are only used during advising-level training, and are computed using either information already available to agents or only require exchange of scalar values (rather than full policy parameters). 
It is sometimes desirable to consider advising under communication constraints, which can be done by deducting a communication cost $c$ from these advising-level rewards for each piece of advice exchanged.

\header{Training Protocol}
Recall LeCTR's two phases are iterated to enable training of increasingly capable advising policies.
In Phase I, task-level learning is conducted using agents' blackbox learning algorithms and latest advising policies.
At the task-level, agents may be independent learners with distinct algorithms.
Advising policies are executed in a decentralized fashion, but their training in Phase II is centralized. 
Our advising policies are trained using the multiagent actor-critic approach of \citeauthor{lowe2017multi} \shortcite{lowe2017multi}. 
Let joint advising-level observations, advising-level actions, and advising-level policies (i.e., `actors') be, respectively, denoted by $\jm{o}=\langle \meta{o}_{\ask}^{i}, \meta{o}_{\ask}^{j}, \meta{o}_{\teach}^{i}, \meta{o}_{\teach}^{j} \rangle$, $\jm{a}=\langle \meta{a}_{\ask}^{i}, \meta{a}_{\ask}^{j}, \meta{a}_{\teach}^{i}, \meta{a}_{\teach}^{j} \rangle$, and $\jm{\pi}=\langle \meta{\pi}_{\ask}^{i}, \meta{\pi}_{\ask}^{j}, \meta{\pi}_{\teach}^{i}, \meta{\pi}_{\teach}^{j} \rangle$, with $\jm{\theta}$ parameterizing $\jm{\pi}$.
To induce $\jm{\pi}$ to learn to teach both agents $i$ and $j$, we use a centralized action-value function (i.e., `critic') with advising-level reward $\meta{r} = \meta{r}^{i}_{\teach} + \meta{r}^{j}_{\teach}$.
Critic $\meta{Q}(\jm{o},\jm{a};\jm{\theta})$ is trained by minimizing loss,
\begin{equation}\label{eq:metacritic_loss}
\thickmuskip=0mu
\!\mathcal{L}(\jm{\theta})=\mathbb{E}_{\jm{o},\jm{a},\meta{r},\jm{o}'\sim \meta{\mathcal{M}}}[(\meta{r}+\gamma\meta{Q}(\jm{o}',\jm{a}';\!\jm{\theta}) - \meta{Q}(\jm{o},\jm{a};\!\jm{\theta})\!)^2]\Big|_{\jm{a}'=\jm{\pi}(\jm{o}')},
\end{equation}
where $\jm{a}'=\jm{\pi}(\jm{o}')$ are next advising actions computed using the advising policies, and $\meta{\mathcal{M}}$ denotes advising-level replay buffer \cite{mnih2015human}.
The policy gradient theorem \cite{sutton2000policy} is invoked on objective $J(\jm{\theta}) = \mathbb{E}[\sum_{t=k}^{T}\meta{\gamma}^{t-k}\meta{r}_t]$ to update advising policies using gradients,
\begin{align}\label{eq:metaactor_grad}
	\nabla_{\jm{\theta}}J(\jm{\theta})
	&=\mathbb{E}_{\jm{o},\jm{a} \sim \meta{\mathcal{M}}} \big[ \nabla_{\jm{\theta}}\log\jm{\pi}(\jm{a}|\jm{o})\meta{Q}(\jm{o},\jm{a};\jm{\theta})\big]
	\\
	\!\!&=\!\mathbb{E}_{\jm{o},\jm{a} \sim \meta{\mathcal{M}}} \big[	\sum_{\mathclap{\substack{\alpha\in\{i,j\}\\\rho\in\{\ask,\teach\}}}} \!\nabla_{\meta{\theta}^{\alpha}_{\rho}} \!\log\meta{\pi}^{\alpha}_{\rho}(\meta{a}^{\alpha}_{\rho}|\meta{o}^{\alpha}_{\rho})\!\nabla_{\meta{a}_{\rho}^{\alpha}}\meta{Q}(\jm{o},\jm{a};\jm{\theta})\big]\!\nonumber,
\end{align}
where $\meta{\pi}^{\alpha}_{\rho}$ is agent $\alpha$'s policy in role $\rho$. 

During training, the advising feedback nonstationarities mentioned earlier are handled as follows: 
in Phase I, task-level policies are trained online (i.e., no replay memory is used so impact of advice on task-level policies is immediately observed by agents);
in Phase II, centralized advising-level learning reduces nonstationarities due to teammate learning, and reservoir sampling is used to further reduce advising reward nonstationarities (see supplementary material for details).
Our overall approach stabilizes advising-level learning.

\header{Advising $n$ Agents}
In the $n$ agent case, students must also decide how to fuse advice from multiple teachers.
This is a complex problem requiring full investigation in future work; 
feasible ideas include using majority voting for advice fusion (as in \citeauthor{da2017simultaneously} \shortcite{da2017simultaneously}), or asking a specific agent for advice by learning a `teacher score' modulated based on teacher knowledge/previous teaching experiences.
\section{Evaluation} \label{sec:evaluation}
\begin{figure}[t!]
	\centering
	
	\begin{subfigure}[t]{0.32\linewidth}
		\centering
		{\renewcommand{\arraystretch}{1.1} 
			\raisebox{0.3cm}{
				\setlength\tabcolsep{4pt}
				\begin{tabular}[b]{cc|cc}
					\multicolumn{2}{c}{} & \multicolumn{2}{c}{Agent $j$}\\
					\multicolumn{1}{c}{} &  & $a_1$  & $a_2$ \\\cline{2-4}
					\parbox[t]{2mm}{\multirow{2}{*}{\rotatebox[origin=c]{90}{Agent $i$}}} & $a_1$ & $0$ & $1$ \\
					& $a_2$ & $0.1$ & $0$ \vspace{0pt}
				\end{tabular}
			}
			\caption{Repeated game payoffs. Each agent has 2 actions ($a_1,a_2$).}
			\label{fig:domain_repeated_game}
		}
	\end{subfigure}
	\hfill
	\begin{subfigure}[t]{0.65\linewidth}
		\includegraphics[width=1\linewidth]{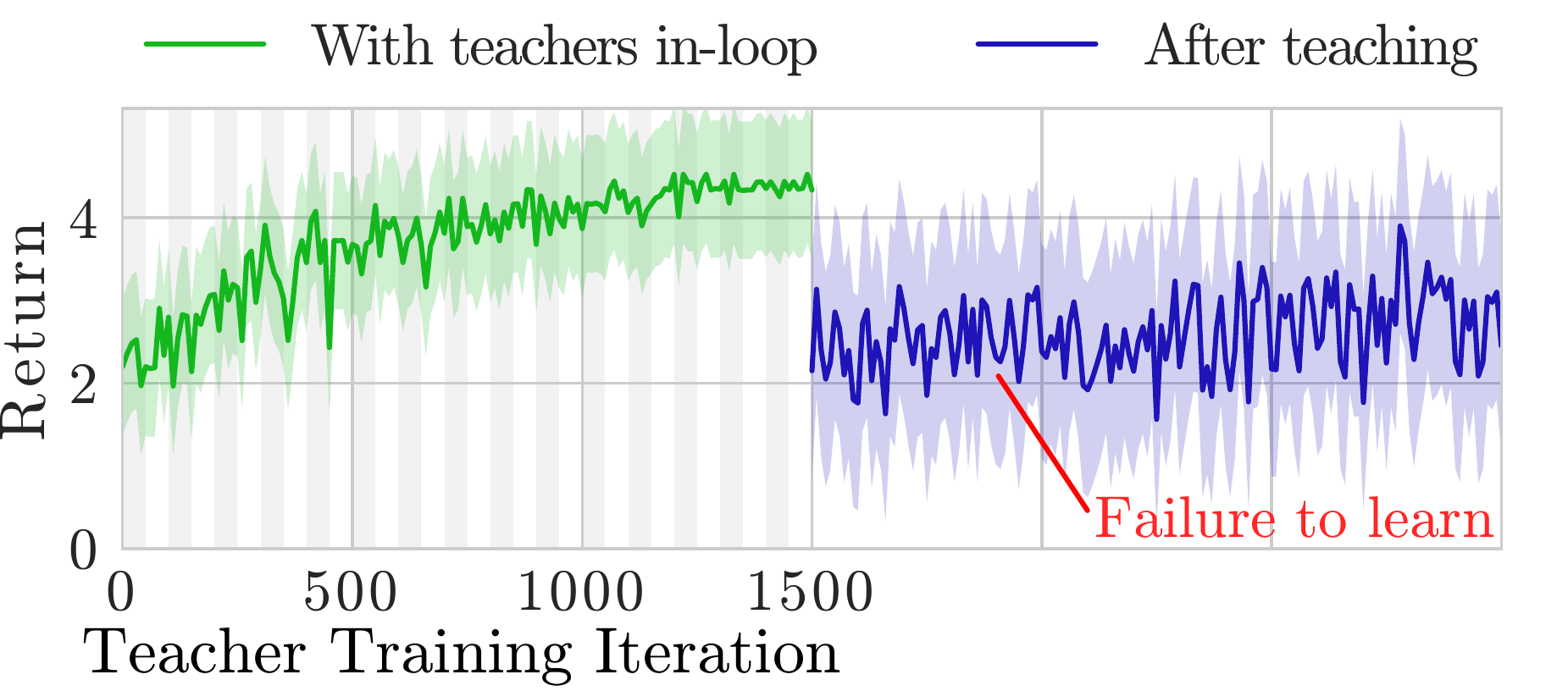}
		\caption{Counterexample showing poor advising reward choice $\meta{r}_{\teach}=r$. LeCTR Phase I and II iterations are shown as background bands.}
		\label{fig:repeated_game_metareward_is_r}
	\end{subfigure}
	\caption{Repeated game. (b) shows a counterexample where using $\meta{r}_{\teach} = r$ yields poor advising, as teachers learn to advise actions that maximize reward (left half, green), but do not actually improve student task-level learning (right half, blue).}
\end{figure}

\begin{figure}[t]
	\centering
	\begin{subfigure}[t]{0.48\linewidth}
		\raisebox{0.7cm}{
			\includegraphics[width=1\linewidth]{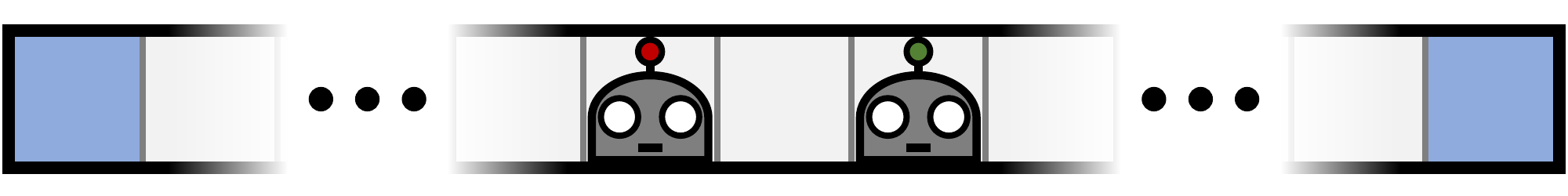}
		}
		\caption{Hallway domain overview.}
		\label{fig:domain_hallway_game}
	\end{subfigure}
	\hfill
	\begin{subfigure}[t]{0.48\linewidth}
		\includegraphics[width=1\linewidth]{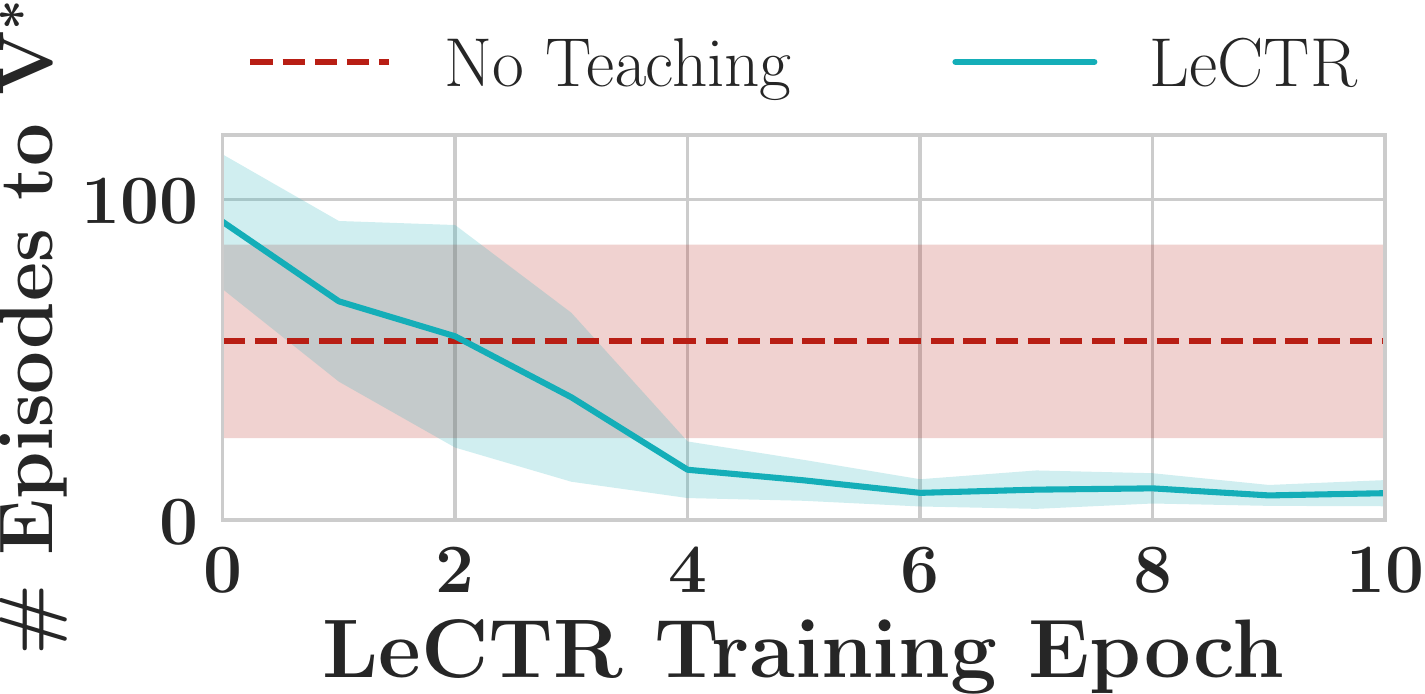}
		\caption{Simult. learning \& teaching}
		\label{fig:hallway_informed_exploration}
	\end{subfigure}
	
	\caption{Hallway game. (a) Agents receive $+1$ reward by navigating to opposite states in $17$-grid hallway. (b) LeCTR accelerates learning \& teaching compared to no-teaching.}
\end{figure}

\begin{figure}[t]
	\centering
	\begin{subfigure}[t]{0.44\linewidth}
		\raisebox{4pt}{
			\includegraphics[width=1\linewidth]{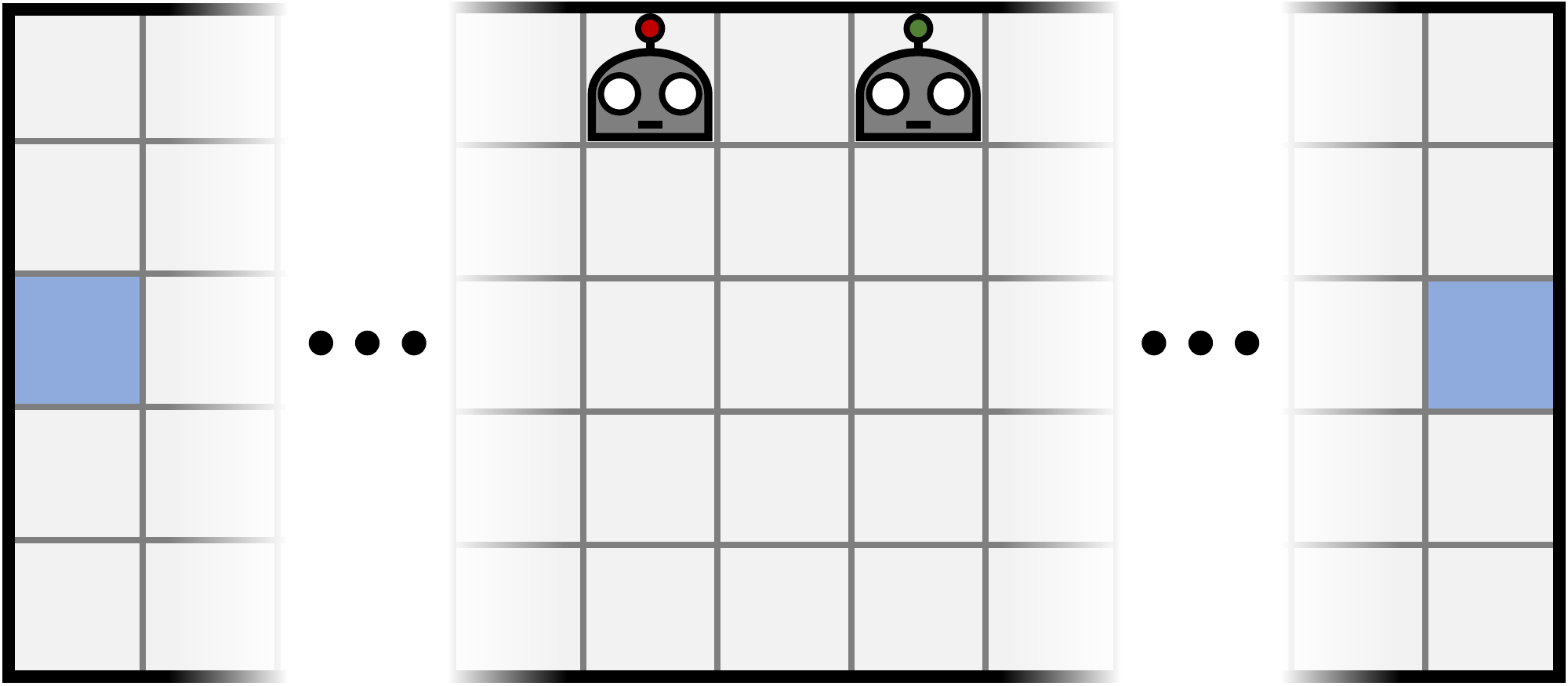}
		}
		\caption{Room domain overview.}
		\label{fig:domain_room_game}
	\end{subfigure}
	\hfill
	\begin{subfigure}[t]{0.52\linewidth}
		\includegraphics[width=1\linewidth]{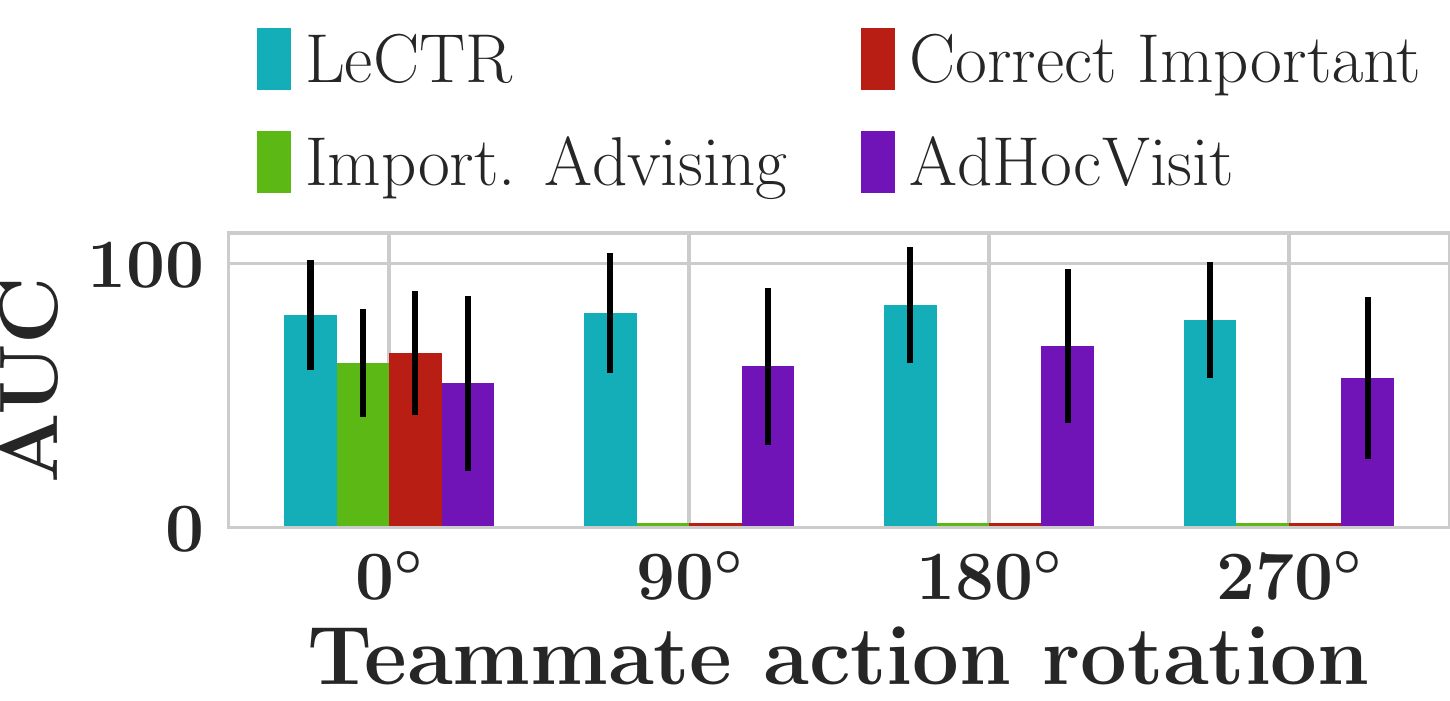}
		\caption{Teaching heterogeneous agents.}
		\label{fig:room_heterogeneous_actions}
	\end{subfigure}
	\caption{Room game. (a) Agents receive $+1$ reward by navigating to opposite goals in $17\! \times\! 5$ grid. (b) LeCTR outperforms prior approaches when agents are heterogeneous.}
\end{figure}

We conduct empirical evaluations on a sequence of increasingly challenging domains involving two agents.
In the `Repeated' game domain, agents coordinate to maximize the payoffs in \cref{fig:domain_repeated_game} over $5$ timesteps.
In `Hallway' (see \cref{fig:domain_hallway_game}), agents only observe their own positions and receive $+1$ reward if they reach opposite goal states; task-level actions are `move left/right', states are agents' joint grid positions.
The higher-dimensional `Room' game (see \cref{fig:domain_room_game}) has the same state/observation/reward structure, but $4$ actions (`move up/right/down/left'). 
Recall student-perspective advising-level actions are to `ask' or `not ask' for advice.
Teacher-perspective actions are to advise an action from the teammate's task-level action space, or to decline advising.

For the Repeated, Hallway, and Room games, respectively, each iteration of LeCTR Phase I consists of $50$, $100$, and $150$ task-level learning iterations.
Our task-level agents are independent Q-learners with tabular policies for the Repeated game and tile-coded policies \cite{sutton1998reinforcement} for the other games.
Advising policies are neural networks with internal rectified linear unit activations. 
Refer to the supplementary material for hyperparameters.
The advising-level learning nature of our problem makes these domains challenging, despite their visual simplicity; 
their complexity is comparable to domains tested in recent MARL works that learn over multiagent learning processes \cite{foerster2018learning}, which also consider two agent repeated/gridworld games.

\header{Counterexample demonstrating \cref{remark:meta_reward_equals_r_case}}
\cref{fig:repeated_game_metareward_is_r} shows results given a poor choice of advising-level reward, $\meta{r}_{\teach}=r$, in the Repeated game.
The left plot (in green) shows task-level return received due to both local policy actions \emph{and} advised actions, which increases as teachers learn.
However, in the right plot (blue) we evaluate how well task-level policies perform by themselves, after they have been trained using the final advising-level policies. 
The poor performance of the resulting task-level policies indicates that advising policies learned to maximize their own rewards $\meta{r}_{\teach}=r$ by always advising optimal actions to students, thereby disregarding whether task-level policies actually learn.
No exploratory actions are advised, causing poor task-level performance after advising.
This counterexample demonstrates that advising-level rewards that reflect student learning progress, rather than task-level reward $r$, are critical for useful advising.

\begin{table*}[t!]
	\centering
	\caption{$\bar{V}$ and Area under the Curve (AUC) for teaching algorithms. Best results in bold (computed via a $t$-test with $p < 0.05$). Independent Q-learning correspond to the no-teaching case. $^\dagger$Final version LeCTR uses the VEG advising-level reward. }
	\label{table:results_comparison}
	\renewrobustcmd{\bfseries}{\fontseries{b}\selectfont}
	\sisetup{detect-weight,mode=text,group-minimum-digits = 4}
	{
		\tabcolsep=0.15cm
		\begin{tabular}[t]{l
				S[separate-uncertainty,table-figures-uncertainty=1,table-format=2.2(2)]
				S[separate-uncertainty,table-figures-uncertainty=1,table-format=3(3)]
				S[separate-uncertainty,table-figures-uncertainty=1,table-format=2.2(2)]
				S[separate-uncertainty,table-figures-uncertainty=1,table-format=3(3)]
				S[separate-uncertainty,table-figures-uncertainty=1,table-format=2.2(2)]
				S[separate-uncertainty,table-figures-uncertainty=1,table-format=3(3)]
			}
			\toprule
			Algorithm                                                &        \multicolumn{2}{c}{Repeated Game}        &       \multicolumn{2}{c}{Hallway Game}       &         \multicolumn{2}{c}{Room Game}         \\
			\cmidrule(lr){2-3} \cmidrule(lr){4-5} \cmidrule(lr){6-7} & $\bar{V}$               & AUC                   & $\bar{V}$               & AUC                & $\bar{V}$               & AUC                 \\ \midrule
			Independent Q-learning (No Teaching)                                           &  2.75 \pm 2.12 & 272 \pm 210           & 0.56 \pm 0.35 & 36 \pm 24          & 0.42 \pm 0.33 & 22 \pm 25          \\
			Ask Important \cite{amir2016interactive}                                           & 1.74 \pm 1.89           & 178 \pm 181           & 0.53 \pm 0.36           & 39 \pm 27          & 0.00 \pm 0.00           & \num{0(0)}          \\
			Ask Uncertain \cite{clouse1996integrating}                                           & 1.74 \pm 1.89           & 170 \pm 184           & 0.00 \pm 0.0            & \num{0(0)}         & 0.00 \pm 0.00           & \num{0(0)}          \\
			Early Advising \cite{torrey2013teaching}                                          & 0.45 \pm 0.00           & 45 \pm 1              & 0.00 \pm 0.0            & \num{0(0)}         & 0.00 \pm 0.00           & \num{0(0)}          \\
			Import. Advising \cite{torrey2013teaching}                                     & 0.45 \pm 0.00           & 45 \pm 1              & 0.67 \pm 0.07           & 39 \pm 17          & 0.57 \pm 0.03           & 48 \pm 8            \\
			Early Correcting \cite{amir2016interactive}                                        & 0.45 \pm 0.00           & 45 \pm 1              & 0.00 \pm 0.0            & \num{0(0)}         & 0.00 \pm 0.00           & \num{0(0)}          \\
			Correct Important \cite{torrey2013teaching}                                       & 0.45 \pm 0.00           & 45 \pm 1              & 0.67 \pm 0.07           & 39 \pm 16          & 0.56 \pm 0.00           & 51 \pm 7            \\
			AdHocVisit \cite{da2017simultaneously}                                              & 2.49 \pm 2.04           & 244 \pm 199           & 0.57 \pm 0.34           & 38 \pm 24          & 0.43 \pm 0.33           & 22 \pm 26           \\
			AdHocTD \cite{da2017simultaneously}                                                 & 1.88 \pm 1.94           & 184 \pm 189           & 0.49 \pm 0.37           & 26 \pm 24          & 0.39 \pm 0.33           & 26 \pm 29           \\ 
			LeCTR (with JVG)                                         & \bfseries 4.16 \pm 1.17 & \bfseries 405 \pm 114 & 0.25 \pm 0.37           & 21 \pm 31          & 0.11 \pm 0.27           & 6 \pm 21            \\
			LeCTR (with QTR)                                         & \bfseries 4.52 \pm 0.00 & \bfseries 443 \pm 3   & 0.21 \pm 0.35           & 12 \pm 22          & 0.20 \pm 0.32           & 11 \pm 22           \\
			LeCTR (with TDG)                                         & 3.36 \pm 1.92           & 340 \pm 138           & 0.19 \pm 0.34           & 15 \pm 25          & 0.26 \pm 0.34           & 25 \pm 32           \\
			LeCTR (with LG)                                          & \bfseries 3.88 \pm 1.51 & 375 \pm 132           & 0.13 \pm 0.29           & 13 \pm 22          & 0.37 \pm 0.34           & 30 \pm 32           \\
			LeCTR (with LGG)                                         & \bfseries 4.41 \pm 0.69 & \bfseries 430 \pm 53  & 0.22 \pm 0.35           & 27 \pm 29          & 0.56 \pm 0.27           & 56 \pm 23           \\
			LeCTR$^\dagger$                                          & \bfseries 4.52 \pm 0.00 & \bfseries 443 \pm 3   & \bfseries 0.77 \pm 0.00 & \bfseries 71 \pm 3 & \bfseries 0.68 \pm 0.07 & \bfseries 79 \pm 16 \\ \bottomrule
		\end{tabular}
	}
\end{table*}

\header{Comparisons to existing teaching approaches}
\cref{table:results_comparison} shows extensive comparisons of existing heuristics-based teaching approaches, no teaching (independent Q-learning), and LeCTR with all advising rewards introduced in \cref{table:metareward_summary}. 
We use the VEG advising-level reward in the final version of our LeCTR algorithm, but show all advising reward results for completeness.
We report both final task-level performance after teaching, $\bar{V}$, and also area under the task-level learning curve (AUC) as a measure of rate of learning; 
higher values are better for both.
Single-agent approaches requiring an expert teacher are extended to the MARL setting by using teammates' policies (pre-trained to expert level) as each agent's teacher.
In the Repeated game, LeCTR attains best performance in terms of final value and rate of learning (AUC). 
Existing approaches always advise the teaching agent's optimal action to its teammate, resulting in suboptimal returns.
In the Hallway and Room games, approaches that tend to over-advise (e.g., Ask Uncertain, Early Advising, and Early Correcting) perform poorly.
AdHocVisit and AdHocTD fare better, as their probabilistic nature permits agents to take exploratory actions and sometimes learn optimal policies.
Importance Advising and Correct Important heuristics lead agents to suboptimal (distant) goals in Hallway and Room, yet attain positive value due to domain symmetries.

LeCTR outperforms all approaches when using the VEG advising-level reward (\cref{table:results_comparison}).
While the JVG advising-level reward seems an intuitive measure of learning progress due to directly measuring task-level performance, its high variance in situations where the task-level value is sensitive to policy initialization sometimes destabilizes training.
JVG is also expensive to compute, requiring game rollouts after each advice exchange.
LG and TDG perform poorly due to the high variance of task-level losses used to compute them. 
We hypothesize that VEG performs best as its thresholded binary advising-level reward filters the underlying noisy task-level losses for teachers.
A similar result is reported in recent work on teaching of supervised learners, where threshold-based advising-level rewards have good empirical performance \cite{fan2018learning}.
\cref{fig:hallway_informed_exploration} shows improvement of LeCTR's advising policies due to training, measured by the number of task-level episodes needed to converge to the max value reached, $V^*$. 
LeCTR outperforms the rate of learning for the no-teaching case, stabilizing after roughly $4$-$6$ training epochs.

\begin{figure}[t]
	\centering
	\begin{subfigure}[t]{0.48\linewidth}
		\includegraphics[width=1\linewidth]{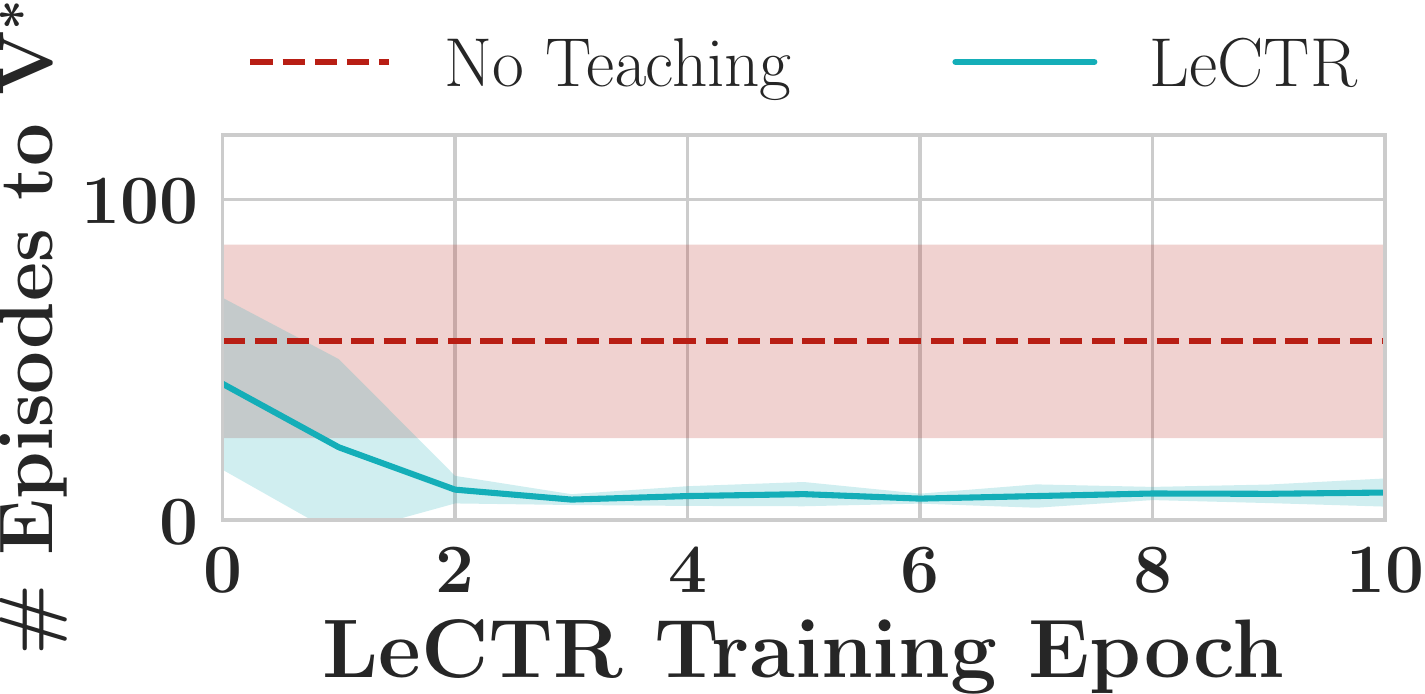}
		\caption{Hallway domain.}
		\label{fig:hallway_transfer}
	\end{subfigure}
	\hfill
	\begin{subfigure}[t]{0.48\linewidth}
		\includegraphics[width=1\linewidth]{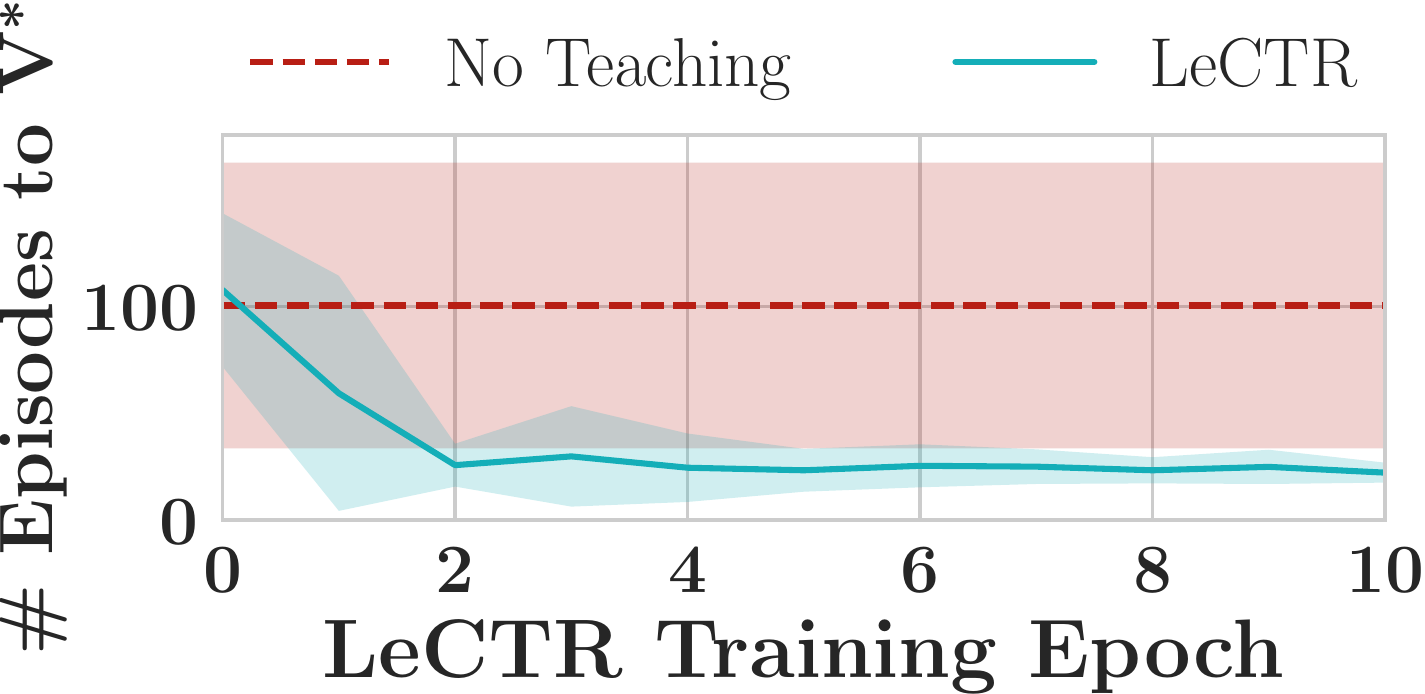}
		\caption{Room domain.}
		\label{fig:room_transfer}
	\end{subfigure}
	\caption{LeCTR accelerates multiagent transfer learning.}
\end{figure}

\begin{figure*}[ht!]
	\centering
	\begin{subfigure}[t]{0.48\linewidth}
		\centering
		\includegraphics[width=0.85\linewidth]{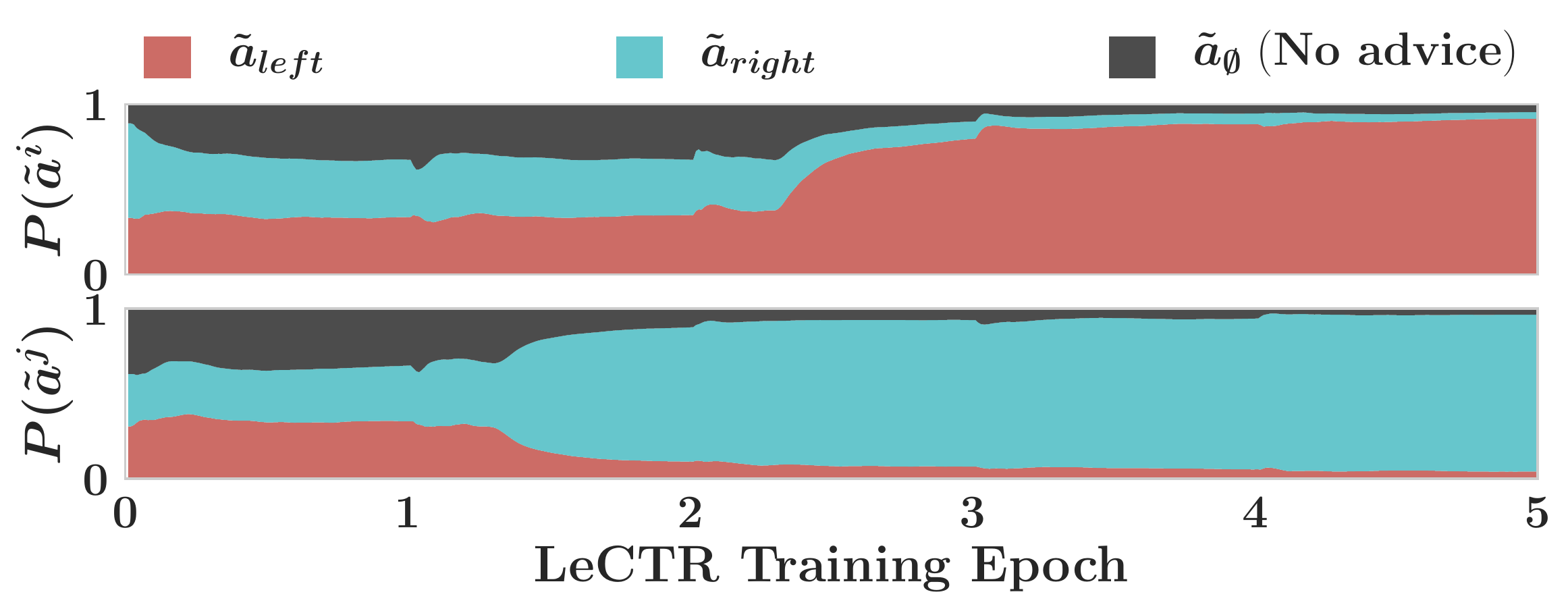}
		\caption{No communication cost, $c=0$, agents advise opposite actions.}
		\label{fig:hallway_game_no_comm_cost}
	\end{subfigure}
	\hfill
	\begin{subfigure}[t]{0.48\linewidth}
		\centering
		\includegraphics[width=0.85\linewidth]{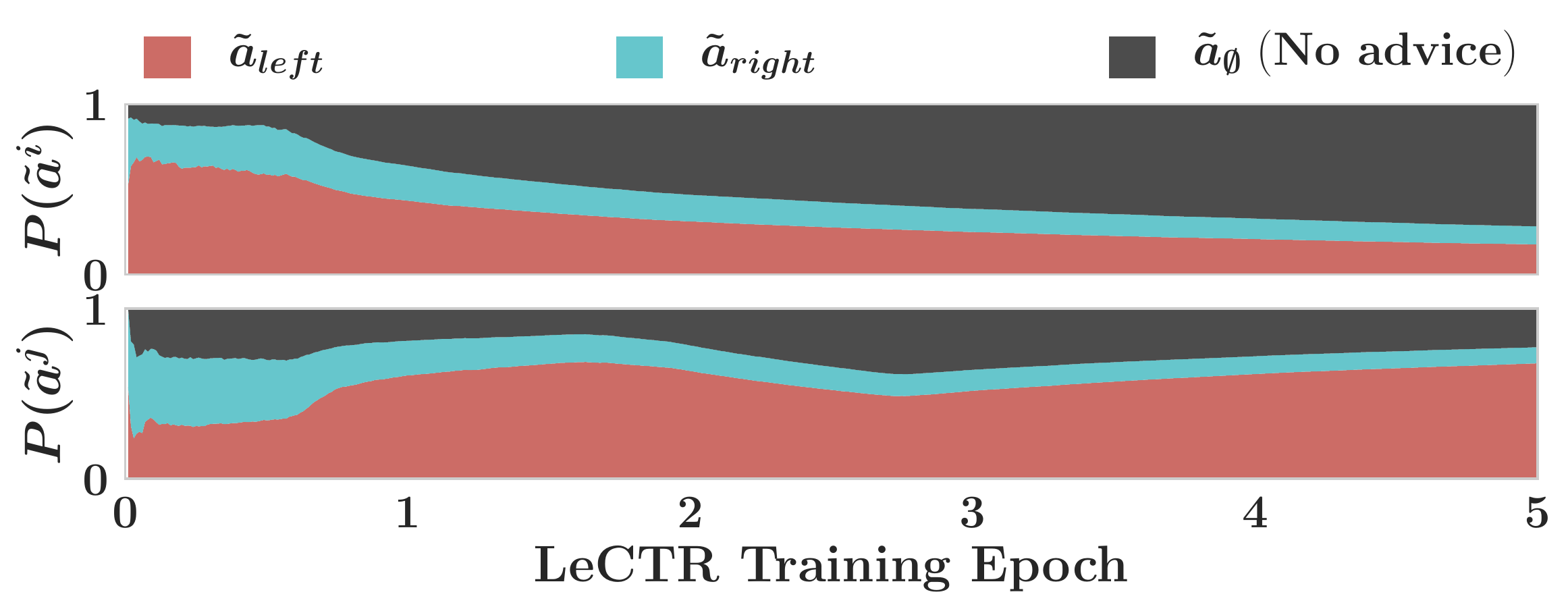}
		\caption{With $c=0.5$, one agent leads \& advises actions opposite its own.}
		\label{fig:hallway_game_comm_cost}
	\end{subfigure}
	\caption{Hallway game, impact of communication cost $c$ on advising policy behaviors. First and second rows show probabilities of action advice, $P(\meta{a}^{i})$ and $P(\meta{a}^{j})$, for agents $i$ and $j$, respectively, as their advising policies are trained using LeCTR.}
	\label{fig:hallway_game_comm_plots}
\end{figure*}

\header{Teaching for transfer learning}
Learning to teach can also be applied to multiagent transfer learning.
We first pre-train task-level policies in the Hallway/Room tasks (denote these $\mathbb{T}_1$), flip agents' initial positions, then train agents to use teammates' $\mathbb{T}_1$ task-level policies to accelerate learning in flipped task $\mathbb{T}_2$.
Results for Hallway and Room are shown in \cref{fig:hallway_transfer,fig:room_transfer}, respectively, where advising accelerates rate of learning using prior task knowledge.
Next, we test \emph{transferability of advising policies} themselves (i.e., use advising policies trained for one task to accelerate learning in a brand new, but related, task).
We {fix} (no longer train) advising policies from the above transfer learning test.
We then consider 2 variants of Room: one with the domain (including initial agent positions) flipped vertically ($\mathbb{T}_3$), and one flipped vertically \emph{and} horizontally ($\mathbb{T}_4$).
We evaluate the fixed advising policies (trained to transfer from $\mathbb{T}_1 \to \mathbb{T}_2$) on transfer from $\mathbb{T}_3 \to \mathbb{T}_4$. 
Learning without advising on $\mathbb{T}_4$ yields AUC $24 \pm 26$, while using the fixed advising policy for transfer $\mathbb{T}_3 \to \mathbb{T}_4$ attains AUC $68 \pm 17$.
Thus, learning is accelerated even when using pre-trained advising policies.
While transfer learning typically involves more significant differences in tasks, these preliminary results motivate future work on applications of advising for MARL transfer learning.

\header{Advising heterogeneous teammates}
We consider heterogeneous variants of the Room game where one agent, $j$, uses rotated versions of its teammate $i$'s action space; 
e.g., for rotation $90^\circ$, agent $i$'s action indices correspond to (\textbf{up}/right/down/left), while $j$'s to (left/\textbf{up}/right/down).
Comparisons of LeCTR and the best-performing existing methods are shown in \cref{fig:room_heterogeneous_actions} for all rotations.
Prior approaches (Importance Advising and Correct Important) work well for homogeneous actions ($0^\circ$ rotation). 
However, they attain 0 AUC for heterogeneous cases, as agents always advise action indices corresponding to their local action spaces, leading teammates to no-reward regions.
AdHocVisit works reasonably well for all rotations, by sometimes permitting agents to explore.
LeCTR attains highest AUC for all rotations.

\header{Effect of communication cost on advice exchange} 
We evaluate impact of communication cost on advising by deducting cost $c$ from advising rewards for each piece of advice exchanged.
\cref{fig:hallway_game_comm_plots} shows a comparison of action advice probabilities for communication costs $c=0$ and $c=0.5$ in the Hallway game.
With no cost ($c=0$ in \cref{fig:hallway_game_no_comm_cost}), agents learn to advise each other opposite actions ($\meta{a}_{\text{left}}$ and $\meta{a}_{\text{right}}$, respectively) in addition to  exploratory actions.
As LeCTR's VEG advising-level rewards are binary ($0$ or $1$), two-way advising nullifies positive advising-level rewards, penalizing excessive advising.
Thus, when $c=0.5$ (\cref{fig:hallway_game_comm_cost}), advising becomes unidirectional: one agent advises opposite exploratory actions of its own, while its teammate tends not to advise.
\section{Related Work} \label{sec:related_work}
Effective diffusion of knowledge has been studied in many fields, including {inverse reinforcement learning} \cite{ng2000algorithms}, {apprenticeship learning} \cite{abbeel2004apprenticeship}, and {learning from demonstration} \cite{argall2009survey}, wherein students discern and emulate key demonstrated behaviors.
Works on {curriculum learning} \cite{bengio2009curriculum} are also related, particularly automated curriculum learning \cite{graves2017automated}.
Though \citeauthor{graves2017automated} focus on single student supervised/unsupervised learning, they highlight interesting measures of learning progress also used here.
Several works meta-learn {active learning policies} for supervised learning \cite{bachman2017learning,fang2017learning,pang2018metalearning,fan2018learning}.
Our work also uses advising-level meta-learning, but in the regime of MARL, where agents must learn to advise teammates without destabilizing coordination.


In action advising, a student executes actions suggested by a teacher, who is typically an expert always advising the optimal action \cite{torrey2013teaching}.
These works typically use state importance value $I(s,\hat{a}) = \max_a Q(s,a) - Q(s,\hat{a})$ to decide when to advise, estimating the performance difference between the student's best action versus intended/worst-case action $\hat{a}$. 
In student-initiated approaches such as Ask Uncertain \cite{clouse1996integrating} and Ask Important \cite{amir2016interactive}, the student decides when to request advice using heuristics based on $I(s,\hat{a})$. 
In teacher-initiated approaches such as Importance Advising \cite{torrey2013teaching}, Early Correcting \cite{amir2016interactive}, and Correct Important \cite{torrey2013teaching}, the teacher decides when to advise by comparing student policy $\pi_{\ask}$ to expert policy $\pi_{\teach}$.
{Q-Teaching} \cite{fachantidis2017learning} learns when to advise by rewarding the teacher $I(s,\hat{a})$ when it advises.
See the supplementary material for details of these approaches.

While most works on information transfer target single-agent settings, several exist for MARL. 
These include imitation learning of expert demonstrations \cite{le2017coordinated}, cooperative inverse reinforcement learning with a human and robot \cite{hadfield2016cooperative}, and transfer to parallel learners in tasks with similar value functions \cite{taylor2013transfer}.
To our knowledge, {AdHocVisit} and {AdHocTD} \cite{da2017simultaneously} are the only action advising methods that do not assume expert teachers;
teaching agents always advise the action they would have locally taken in the student's state, using state visit counts as a heuristic to decide when to exchange advise.
\citeauthor{wang2018efficient} \shortcite{wang2018efficient} uses \citeauthor{da2017simultaneously}'s teaching algorithm with minor changes.


\section{Contribution} \label{sec:contribution}
This work introduced a new paradigm  for learning to teach in cooperative MARL settings. 
Our algorithm, LeCTR, uses agents' task-level learning progress as advising policy feedback, training advisors that improve the rate of learning without harming final performance.
Unlike prior works \cite{torrey2013teaching,taylor2014reinforcement,zimmer2014teacher}, our approach avoids hand-crafted advising policies and does not assume expert teachers.
Due to the many complexities involved, we focused on the pairwise problem, targeting the issues of \emph{when} and \emph{what} to teach. 
A natural avenue for future work is to investigate the $n$-agent setting, extending the ideas presented here where appropriate.

\ifArxivVersion
	\section*{Acknowledgements}
	Research funded by IBM (as part of the MIT-IBM Watson AI Lab initiative) and a Kwanjeong Educational Foundation Fellowship. The authors thank Dr.~Kasra Khosoussi for fruitful discussions early in the paper development process.
\fi

{
	\fontsize{9.5pt}{10.5pt} \selectfont 
	\bibliographystyle{aaai}
	\bibliography{./references}
}

\clearpage
\ifArxivVersion
	\clearpage
	\section{Supplementary Material}
\subsection{Details of Advising-level Rewards}
Recall in Phase II of LeCTR, advising policies are trained to maximize advising-level rewards that should, ideally, reflect the objective of accelerating task-level learning.
Selection of an appropriate advising-level reward is, itself, non-obvious. 
Due to this, we considered a variety of advising-level rewards, each corresponding to a different measure of task-level learning after the student executes an action advice and uses it to update its task-level policy.
Advising-level rewards $\meta{r}_T^j$ below are detailed for the case where agent $i$ is student and agent $j$ teacher (i.e., flip the indices for the reverse case).
Recall that the shared reward used to jointly train all advising policies is $\meta{r} = \meta{r}_T^{i} + \meta{r}_T^{j}$.

\begin{itemize}
	\item \textbf{Joint Value Gain (JVG):} Let $\joint{\theta}_{t}$ and $\joint{\theta}_{t+1}$, respectively, denote agents' joint task-level policy parameters before and after learning from an experience resulting from action advise.
	The JVG advising-level reward measures improvement in task-level value due to advising, such that,
	\begin{equation}
		\meta{r}_{\teach}^{j} = V(s;\joint{\theta}_{t+1})-V(s;\joint{\theta}_{t}).
	\end{equation}
	This is, perhaps, the most intuitive choice of advising-level reward, as it directly measures the gain in task-level performance due to advising.
	However, the JVG reward has high variance in situations where the task-level value is sensitive to policy initialization, which sometimes destabilizes training.
	Moreover, the JVG reward requires a full evaluation of task-level performance after each advising step, which can be expensive due to the game rollouts required.
	
	\item \textbf{Q-Teaching Reward (QTR):} The QTR advising-level reward extends Q-Teaching \cite{fachantidis2017learning} to MARL by using 
	\begin{equation}
		\meta{r}_{\teach}^{j} = I_{\teach}(o^{i},a^{i};h^i) = \max_a Q_{\teach}(\!o^i,a;\!h^i) - Q_{\teach}(\!o^i,a^i;\!h^i),
	\end{equation}
	each time advising occurs. 
	The motivating intuition for QTR is that teacher $j$ should have higher probability of advising when they estimate that the student's intended action, $a^i$, can be outperformed by a different action (the $\arg\max$ action).
	
	\item \textbf{TD Gain (TDG):} For temporal difference (TD) learners, the TDG advising-level reward measures improvement of student $i$'s task-level TD error due to advising, 
	\begin{equation}
		\meta{r}_{\teach}^{j} = |\delta_{t}^{i}|-|\delta_{t+1}^{i}|,
	\end{equation}
	where $\delta_t^i$ is $i$'s TD error at timestep $t$. 
	For example, if agents are independent Q-learners at the task-level, then,
	\begin{equation}
		\delta = r + \max_{a'} Q(o',a';\theta^{i},h'^{i}) - Q(o,a;\theta^{i},h^{i}).
	\end{equation}
	The motivating intuition for the TDG advising-level reward is that actions that are anticipated to reduce student $i$'s task-level TD error should be advised by the teacher $j$.
		
	\item \textbf{Loss Gain (LG):} The LG advising-level reward applies to many loss-based algorithms, measuring improvement of the task-level loss function used by student learner $i$, 
	\begin{equation}
		\meta{r}_{\teach}^{j} = \mathcal{L}(\theta_{t}^{i})-\mathcal{L}(\theta_{t+1}^{i}).
	\end{equation}
	For example, if agents are independent Q-learners using parameterized task-level policies at the task-level, then 
	\begin{equation}
		\mathcal{L}(\theta^{i})=[r+\gamma\max_{a'} Q(o',a';\theta^{i},h'^{i}) - Q(o,a;\theta^{i},h^{i})]^2.
	\end{equation}
	The motivating intuition for the LG reward is similar to the TDG, in that teachers should advise actions they anticipate to decrease student's task-level loss function.
	
	\item \textbf{Loss Gradient Gain (LGG):} The LGG advising-level reward is an extension of the gradient prediction gain \cite{graves2017automated}, which measures the magnitude of student parameter updates due to teaching,
	\begin{equation}
		\meta{r}_{\teach}^{j} = ||\nabla_{\theta^{i}}\mathcal{L}(\theta^{i})||^2_2.
	\end{equation}
	The intuition here is that larger task-level parameter updates may be correlated to learning progress.
	
	\item \textbf{Value Estimation Gain (VEG):} VEG rewards teachers when student's local value function estimates exceed a threshold $\tau$,
	\begin{equation}
		\meta{r}_{\teach}^{j} = \mathds{1}(\hat{V}(\theta^{i})>\tau),
	\end{equation} 
	using $\hat{V}(\theta^{i})=\max_{a^{i}}Q(o^{i},a^{i};\theta^{i},h^{i})$ and indicator function $\mathds{1}(\cdot)$.
	The motivation here is that the student's value function approximation is correlated to its estimated performance as a function of its local experiences.
	A convenient means of choosing $\tau$ is to set it as a fraction of the value estimated when no teaching occurs.
\end{itemize}

\subsection{Details of Heuristics-based Advising Approaches}

Existing works on action advising typically use the state importance value $I_{\rho}(s,\hat{a}) = \max_a Q_{\rho}(s,a) - Q_{\rho}(s,\hat{a})$ to decide when to advise, where $\rho = \inlineask$ for student-initiated advising, $\rho = \inlineteach$ for teacher-initiated advising, $Q_\rho$ is the corresponding action-value function, and $\hat{a}$ is the student's intended action if known (or the worst-case action otherwise). 
$I_{\rho}(s,\hat{a})$ estimates the performance difference of best versus intended student action in state $s$.
The following is a summary of prior advising approaches:
\begin{itemize}
	\item The \textbf{Ask Important} heuristic \cite{amir2016interactive} requests advice whenever $I_{\ask}(s,\hat{a}) \geq k$, where $k$ is a threshold parameter.
	\item \textbf{Ask Uncertain} requests when $I_{\ask}(s,\hat{a}) < k$ \cite{clouse1996integrating}, where $k$ is a threshold parameter.
	\item \textbf{Early Advising} advises until advice budget depletion.
	\item \textbf{Importance Advising} advises when $I_{\teach}(s,a) \geq k$ \cite{torrey2013teaching}, where $k$ is a threshold parameter.
	\item \textbf{Early Correcting} advises when $\pi_{\ask}(s) \neq \pi_{\teach}(s)$ \cite{amir2016interactive}.
	\item \textbf{Correct Important} advises when $I_{\teach}(s) \geq k$ and $\pi_{\ask}(s) \neq \pi_{\teach}(s)$ \cite{torrey2013teaching}, where $k$ is a threshold parameter.
	\item \textbf{Q-Teaching} \cite{fachantidis2017learning} learns when to advise by rewarding the teacher $I_{\teach}(s,\hat{a})$ when advising occurs.
	Constrained by a finite advice budget, Q-Teaching has advising performance similar to Importance Advising, with the advantage of not requiring a tuned threshold $k$.
\end{itemize}
Pairwise combinations of student- and teacher-initiated approaches can be used to constitute a {jointly-initiated} approach \cite{amir2016interactive}, such as ours.
As shown in our experiments, application of single-agent teaching approaches yields poor performance in MARL games.

\subsubsection{Optimal Action Advising}
Note that in the majority of prior approaches, the above heuristics are used to decide \emph{when} to advise.
To address the question of \emph{what} to advise, these works typically assume that teachers have expert-level knowledge and always advise optimal action to students. 
Optimal action advising has a strong empirical track record in single-agent teaching approaches \cite{torrey2013teaching,zimmer2014teacher,amir2016interactive}.
In such settings, the assumed homogeneity of the teacher and student's optimal policies indeed leads optimal action advice to improve student learning (i.e., when the expert teacher's optimal policy is equivalent to student's optimal policy).
In the context of multiagent learning, however, this advising strategy has primarily been applied to games where behavioral homogeneity does not substantially degrade team performance \cite{da2017simultaneously}. 
However, there exist scenarios where multiple agents learn best by exhibiting behavioral diversity (e.g., by exploring distinct regions of the state-action space), or where agents have heterogeneous capabilities/action/observation spaces altogether (e.g., coordination of 2-armed and 3-armed robots, robots with different sensors, etc.).
Use of optimal action advising in cooperative multiagent tasks can lead to suboptimal joint return, particularly when the optimal policies for agents are heterogeneous. We show this empirically in several of our experiments. 

In contrast to earlier optimal advising approaches, our LeCTR algorithm applies to the above settings in addition to the standard homogeneous case; this is due to LeCTR's ability to learn a policy over not only when to advise, but also what to advise.
As shown in our experiments, while existing probabilistic advising strategies (e.g., AdHocTD and AdHocVisit) attain reasonable performance in heterogeneous action settings, they do so passively by permitting students to sometimes explore their local action spaces.
By contrast, LeCTR agents attain even better performance by {actively} learning what to advise within teammates' action spaces; this constitutes a unique strength of our approach.

\subsection{Architecture, Training Details, and Hyperparameters}
At the teaching level, our advising-level critic is parameterized by a 3-layer multilayer perceptron (MLP), consisting of internal rectified linear unit (ReLU) activations, linear output, and 32 hidden units per layer.
Our advising-level actors (advice request/response policies) use a similar parameterization, with the softmax function applied to outputs for discrete advising-level action probabilities.
Recurrent neural networks may also be used in settings where use of advising-level observation histories yields better performance, though we did not find this necessary in our domains.
As in \citeauthor{lowe2017multi} \shortcite{lowe2017multi}, we use the Gumbel-Softmax estimator \cite{jang2016categorical} to compute gradients for the teaching policies over discrete advising-level actions (readers are referred to their paper for additional details).

Policy training is conducted with the Adam optimization algorithm \cite{kingma2014adam}, using a learning rate of $1\mathrm{e}{-3}$.
We use $\gamma = 0.95$ at the task-level and $\meta{\gamma} = 0.99$ at the advising-level level to induce long-horizon teaching policies.
Similar to \citeauthor{graves2017automated} \shortcite{graves2017automated}, we use reservoir sampling to adaptively rescale advising-level rewards with time-varying and non-normalized magnitudes (all except VEG) to the interval $[-1,1]$.
Refer to \citeauthor{graves2017automated} for details on how this is conducted.

\subsection{Experimental Procedures}
In \cref{table:results_comparison}, $\bar{V}$ is computed by running each algorithm until convergence of task-level policies $\joint{\pi}=\langle \pi^{i}, \pi^{j}\rangle$, and computing the mean value obtained by the final joint policy $\joint{\pi}$.
The area under the learning curve (AUC) is computed by intermittently evaluating the resulting task-level policies $\joint{\pi}$ throughout learning; 
while teacher advice is used \emph{during} learning, the AUC is computed by evaluating the resulting $\joint{\pi}$ \emph{after} advising (i.e., in absence of teacher advice actions, such that AUC measures actual student learning progress).
All results and uncertainties are reported using at least 10 independent runs, with most results using over 20 independent runs.
In \cref{table:results_comparison}, best results in bold are computed using a Student's $t$-test with significance level $\alpha = 0.05$.

\newpage
\subsection{Notation}
The following summarizes the notation used throughout the paper.
In general:
superscripts $i$ denote properties for an agent $i$ (e.g., $a^{i}$); 
bold notation denotes joint properties for the team (e.g., $\joint{a} = \langle a^{i}, a^{j} \rangle$);
tilde accents denote properties at the advising-level (e.g., $\meta{a}^{i}$);
and bold characters with tilde accent denote joint advising-level properties (e.g., $\jm{a} = \langle \meta{a}^{i}, \meta{a}^{j} \rangle$).

\begin{table}[h!]
	\centering
	{\renewcommand{\arraystretch}{1.1}
		\begin{tabularx}{\linewidth}{cX}
			\toprule
			Symbol & Definition\\
			\midrule
			$\mathbb{T}$ & Task (a Dec-POMDP)\\
			$\mathbb{L}$ & Task-level learning algorithm\\
			$\joint{\pi}$ & Joint task-level policy\\
			$\pi^{i}$ & Agent $i$'s task-level policy\\
			$\theta^{i}$ & Agent $i$'s task-level policy parameters\\
			$\joint{V}$ & Task-level value\\
			$Q^{i}$ & Agent $i$'s action value function\\
			$\vec{Q}^{i}$ & Agent $i$'s action value vector (i.e., vector of action-values for all actions)\\	    
			$\mathcal{S}$ & State space\\
			$s$ & State\\
			$\mathcal{T}$ & State transition function\\	      
			$\bm{\mathcal{A}}$ & Joint action space\\
			$\joint{a}$ & Joint action\\	    
			$\mathcal{A}^{i}$ & Agent $i$'s action space\\
			$a^{i}$ & Agent $i$'s action\\
			$\bm{\Omega}$ & Joint observation space\\
			$\joint{o}$ & Joint observation\\	    
			$\Omega^{i}$ & Agent $i$'s observation space\\	    
			$o^{i}$ & Agent $i$'s observation\\	    	    
			$\mathcal{O}$ & Observation function\\
			$\mathcal{R}$ & Reward function\\
			$r$ & Task-level reward\\	    	    	    
			$\gamma$ & Discount factor\\
			$\mathcal{M}$ & Task-level experience replay memory\\
			$\delta$ & Task-level temporal difference error\\
			\midrule
			$\rho$ & Agent's advising-level role, where $\rho = \inlineask$ for student role, $\rho = \inlineteach$ for teacher role\\
			$\meta{\pi}^{i}_{\ask}$ & Agent $i$'s advice request policy\\
			$\meta{\pi}^{i}_{\teach}$ & Agent $i$'s advice response policy\\
			$\jm{\pi}$ & Joint advising policy  $\jm{\pi}=\langle \meta{\pi}_{\ask}^{i}, \meta{\pi}_{\ask}^{j}, \meta{\pi}_{\teach}^{i}, \meta{\pi}_{\teach}^{j} \rangle$\\    
			$\jm{\theta}$ & Joint advising-level policy parameters\\    	    
			$\meta{a}^{i}_{\ask}$ & Agent $i$'s advice request action\\
			$\meta{a}_{\emptyset}$ & Special no-advice action\\
			$\meta{a}^{i}_{\teach}$ & Agent $i$'s advice response $\meta{a}^{i}_{\teach} \in \mathcal{A}^{j} \cup \{\meta{a}_{\emptyset}\} $\\
			$\jm{a}$ & Joint advising action $\jm{a}=\langle \meta{a}_{\ask}^{i}, \meta{a}_{\ask}^{j}, \meta{a}_{\teach}^{i}, \meta{a}_{\teach}^{j} \rangle$\\
			$\beta^{i}$ & Agent $i$'s behavioral policy\\
			$\meta{o}^{i}_{\ask}$ & Agent $i$'s student-perspective advising obs.\\
			$\meta{o}^{i}_{\teach}$ & Agent $i$'s teacher-perspective advising obs.\\
			$\jm{o}$ & Joint advising obs. $\jm{o}=\langle \meta{o}_{\ask}^{i}, \meta{o}_{\ask}^{j}, \meta{o}_{\teach}^{i}, \meta{o}_{\teach}^{j} \rangle$\\
			$\meta{r}$ & Advising-level reward $\meta{r}=\meta{r}^{i}_{\teach}+\meta{r}^{j}_{\teach}$\\
			$\meta{\gamma}$ & Advising-level discount factor\\
			$\meta{\mathcal{M}}$ & Advising-level experience replay memory\\	    
			\bottomrule
		\end{tabularx}
	}
\end{table}

\subsection{LeCTR Algorithm}\label{sec:lectr_algorithm_pseudocode}

\begin{algorithm}[h!]
	\caption{Get advising-level observations}
	\label{alg:teaching_meta_obs}
	\begin{algorithmic}[1]
		\Function{GetAdviseObs}{$\joint{o},\joint{\theta}$}
		\For{agents $\alpha \in \{i,j\}$}
		\State Let $-\alpha$ denote $\alpha$'s teammate.
		\State $\meta{o}^{\alpha}_{\ask}\! =\! \langle o^{\alpha}, Q^{\alpha}(o^\alpha;h^{\alpha}) \rangle$
		\State $\meta{o}^{\alpha}_{\teach}\!=\! \langle o^{-\alpha}, Q^{-\alpha}(o^{-\alpha};h^{-\alpha}), Q^{\alpha}(o^{-\alpha};h^{-\alpha}) \rangle$
		\EndFor
		\State\Return $\jm{o} = \langle \meta{o}^{i}_{\ask}, \meta{o}^{j}_{\ask}, \meta{o}^{i}_{\teach}, \meta{o}^{j}_{\teach} \rangle$
		\EndFunction
	\end{algorithmic}  
\end{algorithm}

\begin{algorithm}[h!]
  \caption{LeCTR Algorithm}
  \label{alg:lectr}  
  \begin{algorithmic}[1]
    \For{Phase II episode $\meta{e} = 1$ to $\meta{E}$}
      \State Initialize task-level policy parameters $\joint{\theta}$ 
      \For{Phase I episode ${e} = 1$ to ${E}$}
    	\State $\joint{o} \gets$ initial task-level observation
      	\For{task-level timestep $t = 1$ to $t_{end}$}
			\State $\jm{o} \gets \Call{GetAdviseObs}{\joint{o},\joint{\theta}}$

			\For{agents $\alpha \in \{i,j\}$}
				\State Exchange advice $\meta{a}^\alpha$ via advising policies
				\If{No advising occurred}
					\State Select action $a^\alpha$ via local policy $\pi^{\alpha}$
				\EndIf
			\EndFor

			\State $\joint{a}\gets\langle a^{i}, a^{j} \rangle$, $\jm{a}\gets\langle \meta{a}^{i}, \meta{a}^{j} \rangle$
			
			\State $r,  \joint{o}' \gets$ Execute action $\joint{a}$ in task  
			\State $\theta'^{i} \gets \mathbb{L}^{i}(\theta'^{i}, \langle o^{i}, a^{i}, r,  o'^{i} \rangle)$
			\State $\theta'^{j} \gets \mathbb{L}^{j}(\theta'^{j}, \langle o^{j}, a^{j}, r,  o'^{j} \rangle)$
			\State $\jm{o}' \gets \Call{GetMetaObs}{\joint{o}',\joint{\theta}'}$ 
			\State $\meta{r}^{i}_{\teach},\meta{r}^{j}_{\teach}\gets$ Compute advising-level rewards
           	
           	\State Store $\langle \jm{o}, \jm{a}, \meta{r}= \meta{r}^{i}_{\teach}+\meta{r}^{j}_{\teach} , \jm{o}' \rangle $ in buffer $\meta{\mathcal{M}}$
        \EndFor
      \EndFor
	  \State Update advising-level critic by minimizing loss,
		$$	\mathcal{L}(\jm{\theta}) = \mathbb{E}_{\jm{o},\jm{a},\meta{r},\jm{o}' \sim \meta{\mathcal{M}}}[(\meta{r}+\gamma\meta{Q}(\jm{o}',\jm{a}';\jm{\theta}) - \meta{Q}(\jm{o},\jm{a};\jm{\theta}))^2]\Big|_{\jm{a}'=\jm{\pi}(\jm{o}')}$$
		
      \For{agents $\alpha \in \{i,j\}$}
        \For{roles $\rho \in \{\inlineask, \inlineteach\}$}
          \State Update advising  policy parameters $\meta{\theta}^{\alpha}_{\rho}$ via,
          $$\nabla_{\jm{\theta}}J(\jm{\theta}) 
          	= 	\!\!=\!\mathbb{E}_{\jm{o},\jm{a} \sim \meta{\mathcal{M}}} \big[	\sum_{\mathclap{\substack{\alpha\in\{i,j\}\\\rho\in\{\ask,\teach\}}}} \!\nabla_{\meta{\theta}^{\alpha}_{\rho}} \!\log\meta{\pi}^{\alpha}_{\rho}(\meta{a}^{\alpha}_{\rho}|\meta{o}^{\alpha}_{\rho})\!\nabla_{\meta{a}_{\rho}^{\alpha}}\meta{Q}(\jm{o},\jm{a};\jm{\theta})\big]\!\nonumber,
          	$$
        \EndFor
      \EndFor
	\EndFor
  \end{algorithmic}
\end{algorithm}
\fi

\end{document}